\documentclass[preprint,pof,amsmath,amssymb,aps]{revtex4-1}
\usepackage[abs]{overpic}
\usepackage{graphicx}% Include figure files
\usepackage{dcolumn}% Align table columns on decimal point
\usepackage{bm}% bold math
\usepackage{placeins}
\usepackage{subfig}
\usepackage{color}
\usepackage[normalem]{ulem}
\usepackage{amsmath}
\usepackage{graphicx}
\usepackage{subfig}
\usepackage{amssymb,amsbsy,bm}
\usepackage{placeins}
\usepackage{mathrsfs,amsmath}
\begin{document}

%\preprint{APS/123-QED} 

\title{Developments and difficulties in predicting the relative velocities of inertial particles at the small-scales of turbulence}% Force line breaks with \\
%\thanks{A footnote to the article title}%

\author{Andrew D. Bragg}
\email{andrew.bragg@duke.edu}
\affiliation{Department of Civil \& Environmental Engineering, Duke University, Durham, NC 27708}

\date{\today}% It is always \today, today,
       % but any date may be explicitly specified

\begin{abstract}
In this paper, we use our recently developed theory for the backward-in-time (BIT) relative dispersion of inertial particles in turbulence (Bragg \emph{et al.}, Phys. Fluids 28, 013305, 2016) to develop the theoretical model by Pan \& Padoan (J. Fluid Mech. 661 73, 2010) for inertial particle relative velocities in isotropic turbulence. We focus on the most difficult regime to model, the dissipation range, and find that the modified Pan \& Padoan model (that uses the BIT dispersion theory) can lead to significantly improved predictions for the relative velocities, when compared with Direct Numerical Simulation (DNS) data. However, when the particle separation distance, $r$, is less than the Kolmogorov length scale, $\eta$, the modified model overpredicts the DNS data. We explain how these overpredictions arise from two assumptions in the BIT dispersion theory, that are in general not satisfied when the final separation of the BIT dispersing particles is $<\eta$. We then demonstrate the failure of both the original and modified versions of the Pan \& Padoan model to predict the correct scale-invariant forms for the inertial particle relative velocity structure functions in the dissipation regime. It is shown how this failure, which is also present in other models, is associated with our present inability to correctly predict not only the quantitative, but also the qualitative behavior of the Radial Distribution Function in the dissipation range when $St=\mathcal{O}(1)$.

\end{abstract}

%\pacs{Valid PACS appear here}% PACS, the Physics and Astronomy
               % Classification Scheme.
%\keywords{Suggested keywords}%Use showkeys class option if keyword
               %display desired
\maketitle

%\tableofcontents

%\section{First section}
%
%Article content.
%
%\subsection{A Subsection}

\section{Introduction}
Understanding and predicting the relative motion of inertial particles at the small-scales of turbulence is a problem of fundamental interest with numerous applications. Investigating this relative motion is important because it is physically connected to how the particles mix, disperse, and collide in turbulent flows. Of particular importance is the Relative Velocities (RV) of the inertial particles, and how they depend upon the particle separation. Let $\bm{r}^p(t),\bm{w}^p(t)$ denote the relative separation, and relative velocity vectors, respectively, of two particles. In a turbulent flow, the statistical properties of the RV may be quantified by 
\begin{align}
\varphi(\bm{w},t\vert\bm{r})&\equiv\Big\langle\delta(\bm{w}^p(t)-\bm{w})\Big\rangle_{\bm{r}},\\
\bm{S}^p_{N}(\bm{r},t)&\equiv\int\limits_{\mathbb{R}^3}\bm{w}^N\varphi(\bm{w},t\vert\bm{r})\,d\bm{w}\equiv\Big\langle\Big[\bm{w}^p(t)\Big]^N\Big\rangle_{\bm{r}},
\end{align}
where $\varphi(\bm{w},t\vert\bm{r})$ is the Probability Density Function (PDF) to find the particle pair with relative velocity $\bm{w}$ conditioned on $\bm{r}^p(t)=\bm{r}$, $\langle\cdot\rangle_{\bm{r}}$ denotes an ensemble average conditioned on $\bm{r}^p(t)=\bm{r}$, and $\bm{S}^p_{N}(\bm{r},t)$ is the $N^{th}$-order structure function.

A number of studies have investigated, both theoretically and numerically, how finite particle inertia, quantified by the Stokes number $St$, causes $\varphi(\bm{w},t\vert\bm{r})$ and $\bm{S}^p_{N}(\bm{r},t)$ to deviate from the corresponding forms for fluid particles ($St=0$) \cite{wilkinson05,falkovich07,bec10a,bec11,gustavsson11,gustavsson12,bragg14c,ireland16a}. Most of these studies have concentrated on point-particles that are advected by a Stokes drag force, which applies to particles whose diameter is $\ll\eta$, and whose material density is much greater than that of the fluid in which they are suspended. This is also the system that we shall consider in the present paper.

For monodisperse particles, our understanding of the physical mechanisms controlling $\varphi(\bm{w},t\vert\bm{r})$ and $\bm{S}^p_{N}(\bm{r},t)$ for finite $St$ is now well developed. We refer the reader to \cite{bec10a,bragg14c,ireland16a} for detailed explanations; here we summarize the understanding for $r=\|\bm{r}\|$ in the dissipation regime of turbulence (the focus of the present paper). When $St\ll1$, the dominant mechanism causing deviation from the $St=0$ case is the preferential sampling mechanism. Inertial particles interact with the topology of the turbulent velocity field in such a way that they do not sample the fluid velocity field uniformly (unlike $St=0$ particles), showing a tendancy to preferentially sample regions of the flow where the local velocity gradient tensor is dominated by straining motions. When $St\geq\mathcal{O}(1)$, the particles are affected by their finite memory of the fluid velocity field they have experienced along their path-history. Since the fluid velocity differences that drive $\bm{w}^p(t)$ depend upon separation, then inertial particles at a given separation $\bm{r}$ can be affected by their memory of the fluid velocity differences at larger separations in the past. This leads to a dramatic increase in the RV with increasing $St$, and to the phenomena of ``caustics'' \cite{wilkinson05}, ``the sling effect'' \cite{falkovich07}, and ``random uncorrelated motion'' \cite{ijzermans10}. Finally, if $St$ is sufficiently large, then the filtering effect of particle inertia takes over, which is associated with the inability of highly inertial particles to respond to fluctuations in the underlying turbulent velocity field. When the filtering effect dominates the dynamics, it causes the RV to decrease with increasing $St$.

Although our understanding of the influence of $St$ on $\varphi(\bm{w},t\vert\bm{r})$ and $\bm{S}^p_{N}(\bm{r},t)$ is essentially complete for the particular dynamical system described above, in general, we are unable to accurately predict the effect of $St$ on these quantities. A number of theoretical models have attempted to predict $\bm{S}^p_{N}(\bm{r},t)$ for $N\leq 2$. In \cite{bragg14c} we compared the predictions of some of these models against DNS data and found that in general the model predictions were at best moderately accurate, showing the greatest errors when $St=\mathcal{O}(1)$. The most successful of the models that were tested was that by Pan \& Padoan \cite{pan10}, though it consistently underpredicted the DNS data across the range $St\in(0,3]$. We argued in \cite{bragg14c} that a possible cause of these under-predictions was that in the Pan \& Padoan  model (PPM hereafter), they approximated the backward-in-time (BIT) mean-square separation of the inertial particles by the forward-in-time (FIT) counterpart, and that the BIT was likely faster than the FIT separation. Pan \& Padoan invoked this approximation because the BIT mean-square separation of inertial particles in turbulence had never been studied before, and so they were forced to approximate it by the FIT version, guided by the results in \cite{bec10b}.

Motivated in part by such issues, we recently completed a theoretical and numerical study on the BIT relative dispersion of inertial particles in turbulence \cite{bragg16}. Our results demonstrated that for inertial particles, BIT dispersion could be much faster than FIT dispersion, differing by as much as two orders of magnitude. This clearly has implications for the PPM that approximated the BIT dispersion by the FIT counterpart. The aim of the present paper is to apply the theoretical results from \cite{bragg16} to the PPM, and see if it does in fact improve the model predictions, as was conjectured in \cite{bragg14c}.

The outline of the remainder of the paper is as follows. In \S\ref{MD} we explain how the PPM can be modified to incorporate the BIT dispersion theory from \cite{bragg16}. In \S\ref{Sp2results} we then compare the modified version of the PPM with the original version and with DNS data. In \S\ref{rdfimp} we consider the implications of the results of the PPM and its modified version for predicting the Radial Distribution Function (RDF) of inertial particles in turbulence. Finally, in \S\ref{conc} we draw conclusions to the work and highlight future problems that must be addressed.

\section{Developing the Pan \& Padoan model}\label{MD}
The PPM is derived for monodisperse and bidisperse particles subject to Stokes drag forcing in isotropic turbulence. We shall be concerned with the monodisperse case, for which the equation of motion is
\begin{align}
\ddot{\bm{r}}^p(t)\equiv\dot{\bm{w}}^p(t)=\frac{1}{\tau_p}\Big(\Delta\bm{u}(\bm{x}^p(t),\bm{r}^p(t),t)-\bm{w}^p(t)\Big),\label{reom}
\end{align}
where $\tau_p$ is the particle response time, and $\Delta\bm{u}(\bm{x}^p(t),\bm{r}^p(t),t)$ is the difference in the fluid velocity experienced by a pair of particles located at $\bm{x}^p(t)$ and $\bm{x}^p(t)+\bm{r}^p(t)$ (when $\Delta\bm{u}$ appears in statistical expressions we will drop the $\bm{x}^p(t)$ coordinate since the system we are considering is spatially homogeneous). The solution of (\ref{reom}) for $\bm{w}^p(t)$ may be represented in the integral form
\begin{align}
\bm{w}^p(t)=\dot{\mathcal{G}}(t,0)\bm{w}^p(0)+\frac{1}{\tau_p}\int\limits_0^t\dot{\mathcal{G}}(t,t')\Delta\bm{u}(\bm{x}^p(t'),\bm{r}^p(t'),t')\,dt',\label{wpsol}
\end{align}
where $\dot{\mathcal{G}}(t,t')\equiv (d/dt)\mathcal{G}(t,t')$ and $\mathcal{G}(t,t')\equiv\tau_p[1- e^{-(t-t')/\tau_p}]$. Using (\ref{wpsol}), Pan \& Padoan then construct the exact integral equation governing $\bm{S}^p_{2}(\bm{r},t)$, and finally arrive at a closed equation by applying a series of approximations to the Lagrangian statistics of $\Delta\bm{u}(\bm{x}^p(t'),\bm{r}^p(t'),t')$. 

Solving the integrand in (\ref{wpsol}) requires knowledge of the BIT locations of the particle-pair, since the integrand depends upon $\bm{r}^p(t')$, and $t'\in[0,t]$. Due to the simplifying assumptions made in PPM, the actual BIT statistic requiring closure in the integral equation for $\bm{S}^p_{2}(\bm{r},t)$ is $\langle\|\bm{r}^p(t')\|^2\rangle_{\bm{r}}$. The quantity $\langle\|\bm{r}^p(t')\|^2\rangle_{\bm{r}}$ is the mean-square separation of inertial particles at time $t'\leq t$ (and hence BIT) evaluated along trajectories that satisfy the condition $\bm{r}^p(t)=\bm{r}$. Pan \& Padoan construct their theoretical model for a steady state condition, for which the terminal time $t$ is arbitrary, and the statistics depend only upon $t-t'$. We may then take $t=0$ and introduce $t-t'\to s \in[0,\infty]$. The BIT mean-square separation  appearing in the integral equation for $\bm{S}^p_{2}(\bm{r})$ is then written as $\langle\|\bm{r}^p(-s)\|^2\rangle_{\bm{r}}$ (where now the conditionality is $\bm{r}^p(0)=\bm{r}$).

Partly guided by the results in \cite{bec10b}, Pan \& Padoan approximated $\langle\|\bm{r}^p(-s)\|^2\rangle_{\bm{r}}$ by the following piecewise function
\begin{align}
\Big\langle\|\bm{r}^p(-s)\|^2\Big\rangle_{\bm{r}}\approx
  \begin{cases}
    r^2+\mathrm{tr}[\bm{S}^p_2(\bm{r})]s^2 & \quad \text{for}\,\,s\in[0,s_c)\\
		    \langle\|\bm{r}^p(-s_c)\|^2\rangle_{\bm{r}}+\mathfrak{g}\langle\epsilon\rangle (s-s_c)s^2 & \quad \text{for}\,\,s\in[s_c,s_d)\\
						    L^2+2\mathcal{D} (s-s_d) & \quad \text{for}\,\,s\in[s_d,\infty].\label{r2regimes}
  \end{cases}
\end{align}
In this expression, $s_c\equiv (7/5)\tau_p$, $\mathfrak{g}$ is the BIT Richarson constant, which they take to be $\mathfrak{g}=1$, $\langle\epsilon\rangle$ is the fluid turbulent kinetic energy dissipation rate, $L$ is the integral lengthscale of the flow, $\mathcal{D}\equiv 6 u'u' /\tau_I$ is a large-scale diffusion coefficient, $u'$ is the fluid velocity r.m.s. value, and $\tau_I$ is the fluid integral timescale. The time $s_d$ is defined through $\sqrt{\langle\|\bm{r}^p(-s_d)\|^2\rangle_{\bm{r}}}\equiv L$. Note that the expression in (\ref{r2regimes}) for $s\in[0,s_c)$ involves $\bm{S}^p_2(\bm{r})$, which is in fact the tensor function that the PPM is constructed to predict. The consequence of this is that the PPM integral equation for $\bm{S}^p_{2}(\bm{r})$ must be solved iteratively.

The three regimes in (\ref{r2regimes}) correspond first to ballistic motion, then to Richardson dispersion, and finally to large-scale diffusion, where the mean-square separation grows linearly with time. The specification for $s\in[0,s_c)$ is based upon the results for the FIT dispersion of inertial particles in isotropic turbulence from Bec \emph{et al.}\cite{bec10b}. Therefore, the PPM invokes the approximation $\langle\|\bm{r}^p(-s)\|^2\rangle_{\bm{r}}\approx \langle\|\bm{r}^p(s)\|^2\rangle_{\bm{r}}$ for $s\in[0,s_c)$. We will return to this momentarily. The PPM partially captures the BIT nature of the dispersion in the regime $s\in[s_c,s_d)$ in that it uses the BIT value for $\mathfrak{g}$ instead of the FIT value, though it does not include corrections to Richardson's law that arise due to particle inertia \cite{bec10b,bragg16}. For $s\in[s_d,\infty]$, the the particle separations are greater than $L$, and for homogeneous turbulence, the dispersion is precisely reversible at these scales \cite{bragg16}.

In light of these considerations, there are questions concerning the specification of $\langle\|\bm{r}^p(-s)\|^2\rangle_{\bm{r}}$ in (\ref{r2regimes}) with respect to its form for $s\in[0,s_c)$ and $s\in[s_c,s_d)$. The results of our recent study in \cite{bragg16} showed that for $s\in[0,s_c)$ and $r$ in the dissipation range, the dispersion of inertial particles is strongly irreversible, with $\langle\|\bm{r}^p(-s)\|^2\rangle_{\bm{r}}\gg\langle\|\bm{r}^p(s)\|^2\rangle_{\bm{r}}$ for $s=\mathcal{O}(s_c)$. In \cite{bragg16} we developed a theoretical prediction for $\langle\|\bm{r}^p(-s)\|^2\rangle_{\bm{r}}$ that is valid for $s\in[0,s_c)$, and comparisons with DNS data demonstrated its accuracy. For $r$ in the dissipation regime, the theoretical prediction is 
\begin{align}
\begin{split}
\Big\langle\|\bm{r}^p(-s)\|^2\Big\rangle_{\bm{r}}\approx &\|\bm{r}\|^2+\mathcal{G}^2(-s)\mathrm{tr}[\bm{S}^p_2]+\Big(\mathcal{G}^2(-s)+2s\mathcal{G}(-s)+s^2 \Big)\mathrm{tr}[\bm{S}^f_2]\\
&-2\mathcal{G}(-s)\Big(\mathcal{G}(-s)+s\Big)\sqrt{\mathrm{tr}[\bm{S}^p_2]\mathrm{tr}[\bm{S}^f_2]},\quad s\leq\mathcal{O}(\tau_p),\label{BITclos}
\end{split}
\end{align}
where $\mathcal{G}(-s)\equiv\tau_p(1-e^{s/\tau_p})$ and $\bm{S}^f_2(\bm{r})\equiv \langle\|\Delta\bm{u}(\bm{r},0)\|^2\rangle$. In order to correctly describe the BIT mean-square separation for $s\in[0,s_c)$ in the PPM, we can simply replace $r^2+\mathrm{tr}[\bm{S}^p_2(\bm{r})]s^2$ in (\ref{r2regimes}) with the rhs of (\ref{BITclos}). It is important to note that in doing so we are not introducing any additional quantities or unknowns to the PPM; (\ref{BITclos}) only depends upon quantities that are already present in the PPM.

In \cite{bragg16} we also developed a theoretical prediction that is valid in the same regime as Richardson's law, provided that ${St_r(s)\ll1}$, where $St_r(s)\equiv\tau_p/\tau_r(s)$ and $\tau_r(s)$ is the fluid eddy turnover time based upon the particle separation at time $s$. The prediction is
\begin{align}
	\Big\langle\|\bm{r}^p(-s)\|^2\Big\rangle_{\bm{r}}\approx\mathfrak{g}\langle\epsilon\rangle s^3\Big(1+(\tau_p\mathcal{A}/2)\mathfrak{g}^{-4/3}s^{-1}\ln[s/\tau_p]\Big)+\mathcal{O}\Big(St^2_r(s)\Big),\quad\text{for}\,s\gg\tau_p,\label{r2larges}
\end{align}
where $\mathcal{A}\approx 39.13$. Equation (\ref{r2larges}) predicts that inertia leads to an enhancement of the BIT mean-square separation, and formally reduces to Richarson's law when either $St=0$, or else in the limit $s\to\infty$ if $St>0$.

Since the condition ${St_r(s)\ll1}$ may not in general be satisfied for $s\in[s_c,s_d)$, using (\ref{r2larges}) may lead to errors. We will therefore consider two modified versions the Pan \& Padoan model. The first, which we denote by PPM$^*$, modifies PPM (the original model) by using (\ref{BITclos}) to prescribe $\langle\|\bm{r}^p(-s)\|^2\rangle_{\bm{r}}$ for $s\in[0,s_c)$, but retains the original specification for $s\in[s_c,s_d)$. The second version, which we denote by PPM$^{**}$, modifies PPM both by using (\ref{BITclos}) to prescribe $\langle\|\bm{r}^p(-s)\|^2\rangle_{\bm{r}}$ for $s\in[0,s_c)$, and also uses (\ref{r2larges}) to prescribe $\langle\|\bm{r}^p(-s)\|^2\rangle_{\bm{r}}$ for $s\in[s_c,s_d)$. Note that analogous to the use of Richardson's law in (\ref{r2regimes}), when using (\ref{r2larges}) in PPM$^{**}$ we actually use the modified version
\begin{align}
	\Big\langle\|\bm{r}^p(-s)\|^2\Big\rangle_{\bm{r}}\approx\Big\langle\|\bm{r}^p(-s_c)\|^2\Big\rangle_{\bm{r}}+\mathfrak{g}\langle\epsilon\rangle s^2(s-s_c)\Big(1+(\tau_p\mathcal{A}/2)\mathfrak{g}^{-4/3}s^{-1}\ln[s/\tau_p]\Big),\quad s\in[s_c,s_d),\label{r2larges2}
\end{align}
to ensure an appropriate transition from the $s\in[0,s_c)$ to the $s\in[s_c,s_d)$ regime.
\section{Results for $\bm{S}^p_2$}\label{Sp2results}
In this section, we compare the predictions from PPM, PPM$^*$ and PPM$^{**}$ with DNS data. The DNS data is for statistically stationary, homogeneous, isotropic, particle-laden turbulence at $Re_\lambda=398$. The data comes from the same set presented in detail in \cite{ireland16a}, to which we therefore refer the reader for a detailed account, and to \cite{ireland13} for a detailed account of the DNS methodology.

\begin{figure}[ht]
\centering
\vspace{-7mm}
\subfloat[]
{\begin{overpic}
[trim = 20mm 70mm 24mm 60mm,scale=0.45,clip,tics=20]{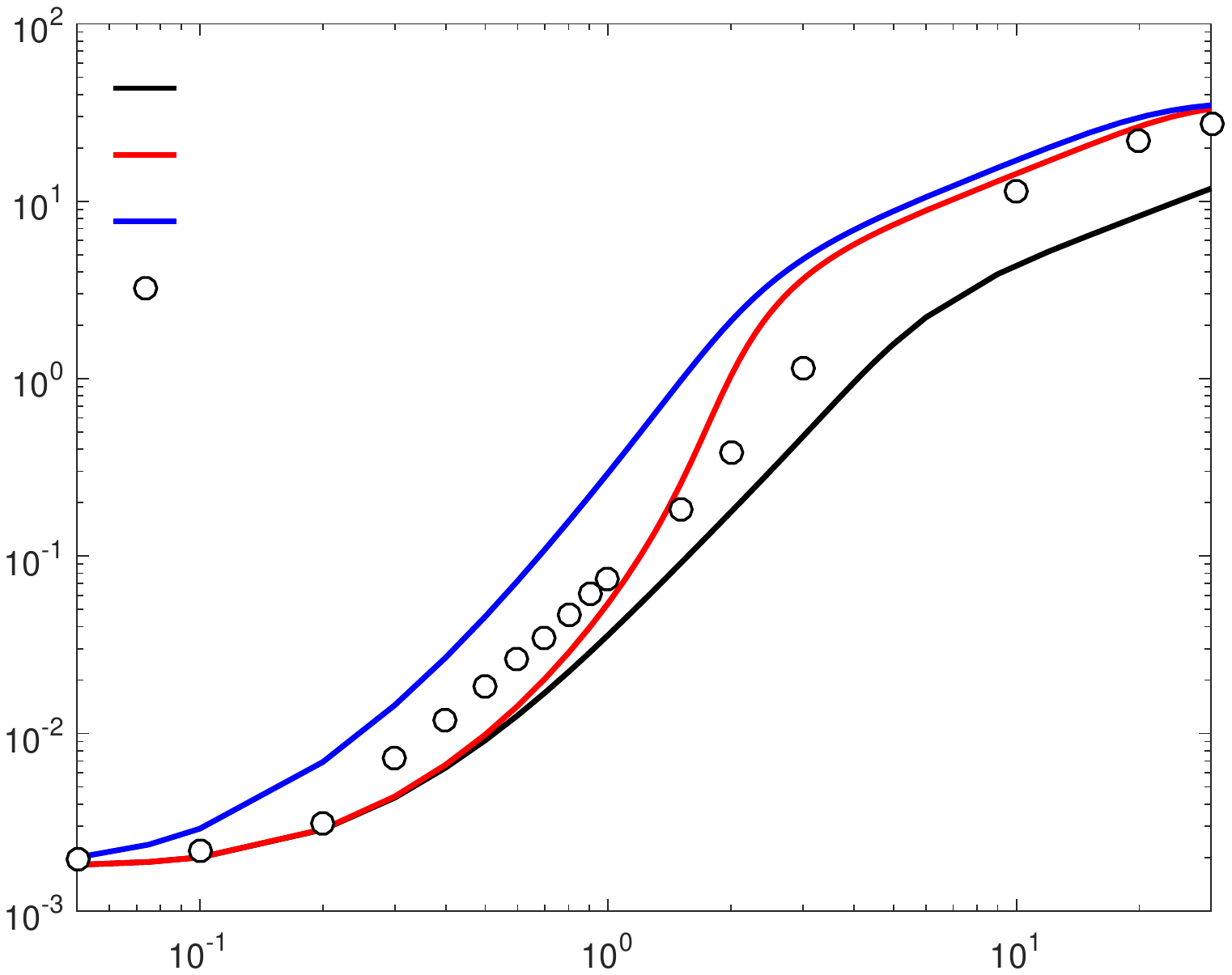}
\put(105,-2){$St$}
\put(0,75){\rotatebox{90}{$S^p_{2\parallel}/u_\eta^2$}}
\put(45,149){\small{PPM}}
\put(45,139){\small{PPM$^*$}}
\put(45,129){\small{PPM$^{**}$}}
\put(45,119){\small{DNS}}
\end{overpic}}
\subfloat[]
{\begin{overpic}
[trim = 20mm 70mm 24mm 60mm,scale=0.45,clip,tics=20]{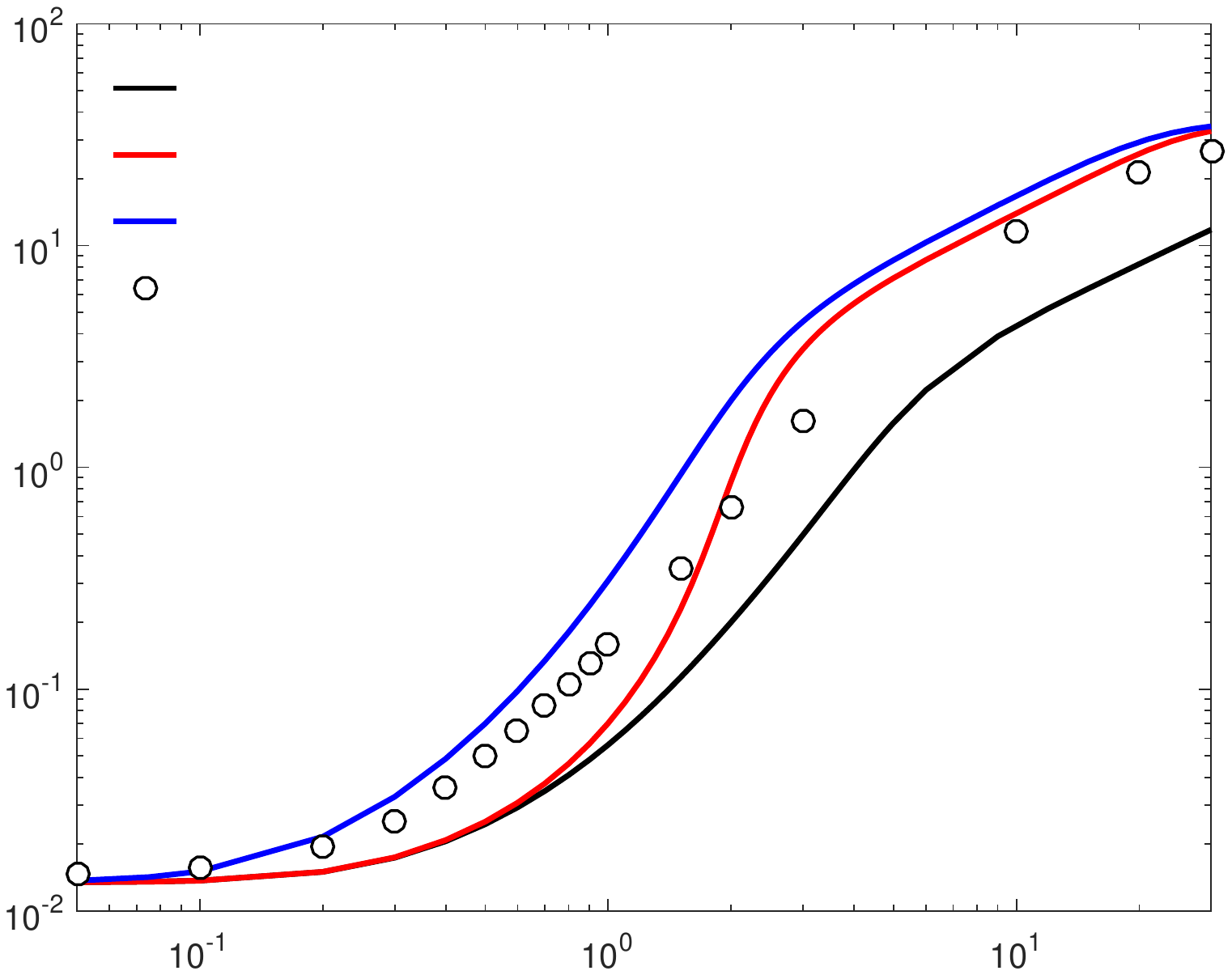}
\put(105,-2){$St$}
\put(0,75){\rotatebox{90}{$S^p_{2\parallel}/u_\eta^2$}}
\put(45,149){\small{PPM}}
\put(45,139){\small{PPM$^*$}}
\put(45,129){\small{PPM$^{**}$}}
\put(45,119){\small{DNS}}
\end{overpic}}\\
\vspace{-10mm}
\subfloat[]
{\begin{overpic}
[trim = 20mm 70mm 24mm 60mm,scale=0.45,clip,tics=20]{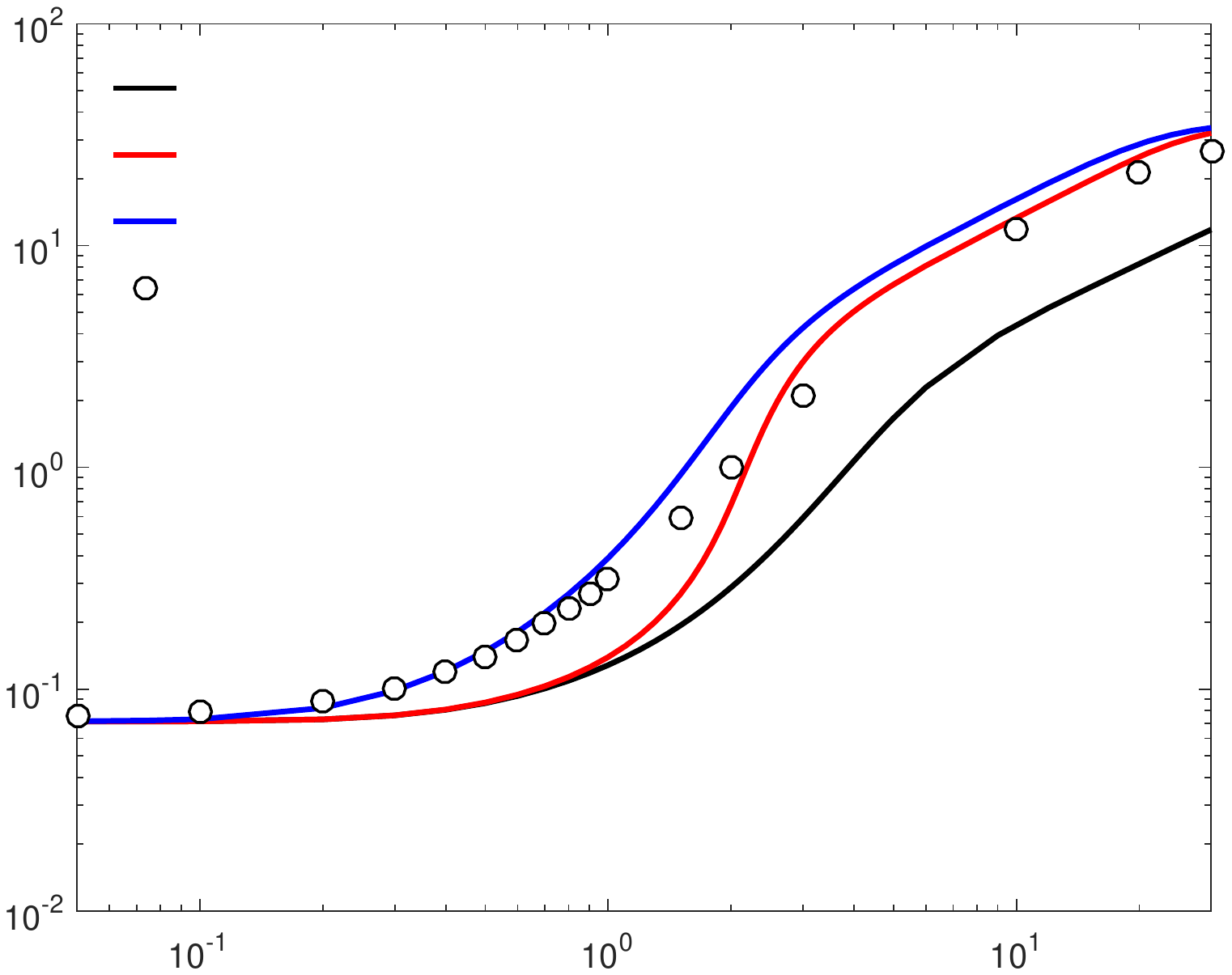}
\put(105,-2){$St$}
\put(0,75){\rotatebox{90}{$S^p_{2\parallel}/u_\eta^2$}}
\put(45,149){\small{PPM}}
\put(45,139){\small{PPM$^*$}}
\put(45,129){\small{PPM$^{**}$}}
\put(45,119){\small{DNS}}
\end{overpic}}
\subfloat[]
{\begin{overpic}
[trim = 20mm 70mm 24mm 60mm,scale=0.45,clip,tics=20]{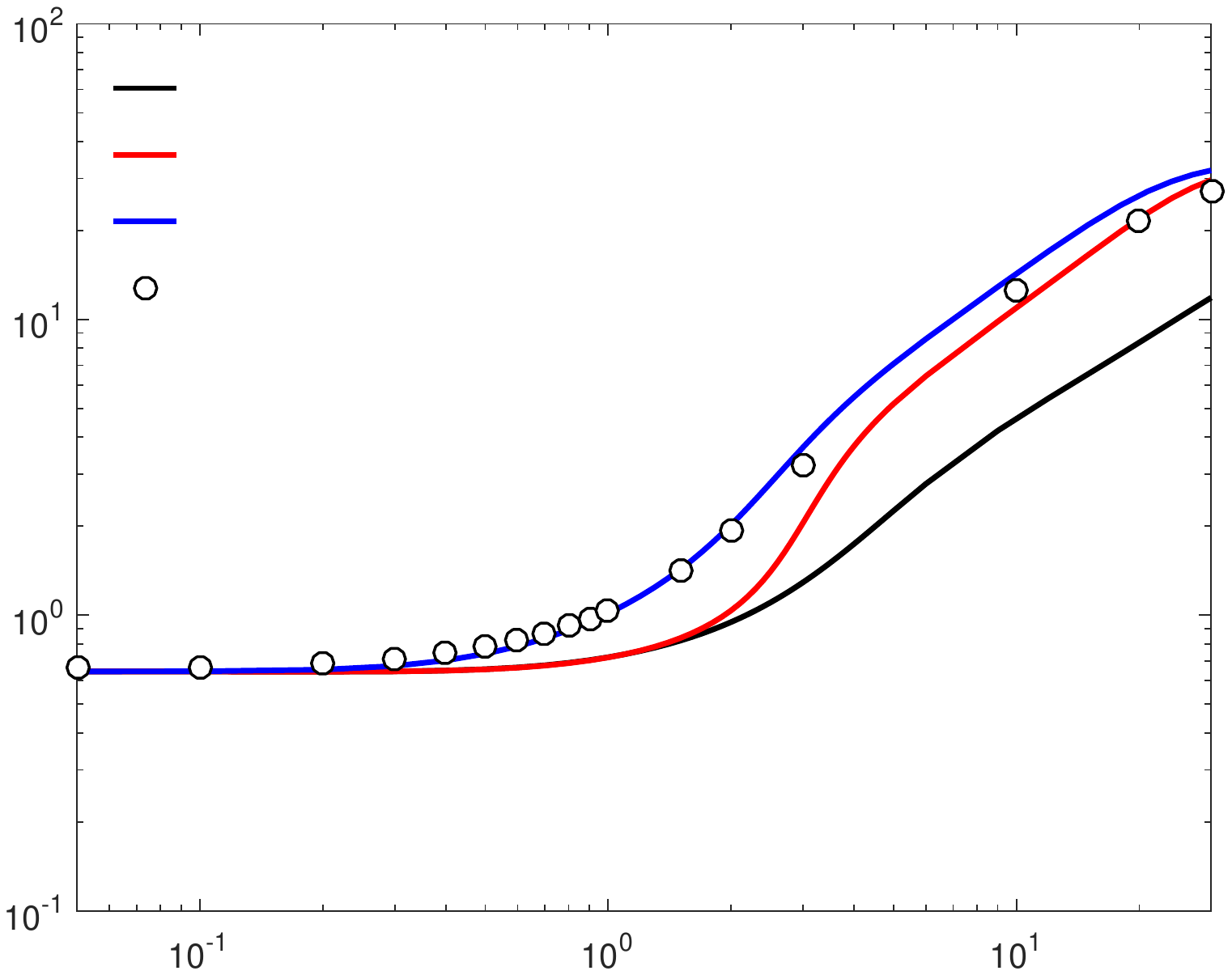}
\put(105,-2){$St$}
\put(0,75){\rotatebox{90}{$S^p_{2\parallel}/u_\eta^2$}}
\put(45,149){\small{PPM}}
\put(45,139){\small{PPM$^*$}}
\put(45,129){\small{PPM$^{**}$}}
\put(45,119){\small{DNS}}
\end{overpic}}\\
\vspace{-10mm}
\subfloat[]
{\begin{overpic}
[trim = 20mm 70mm 24mm 60mm,scale=0.45,clip,tics=20]{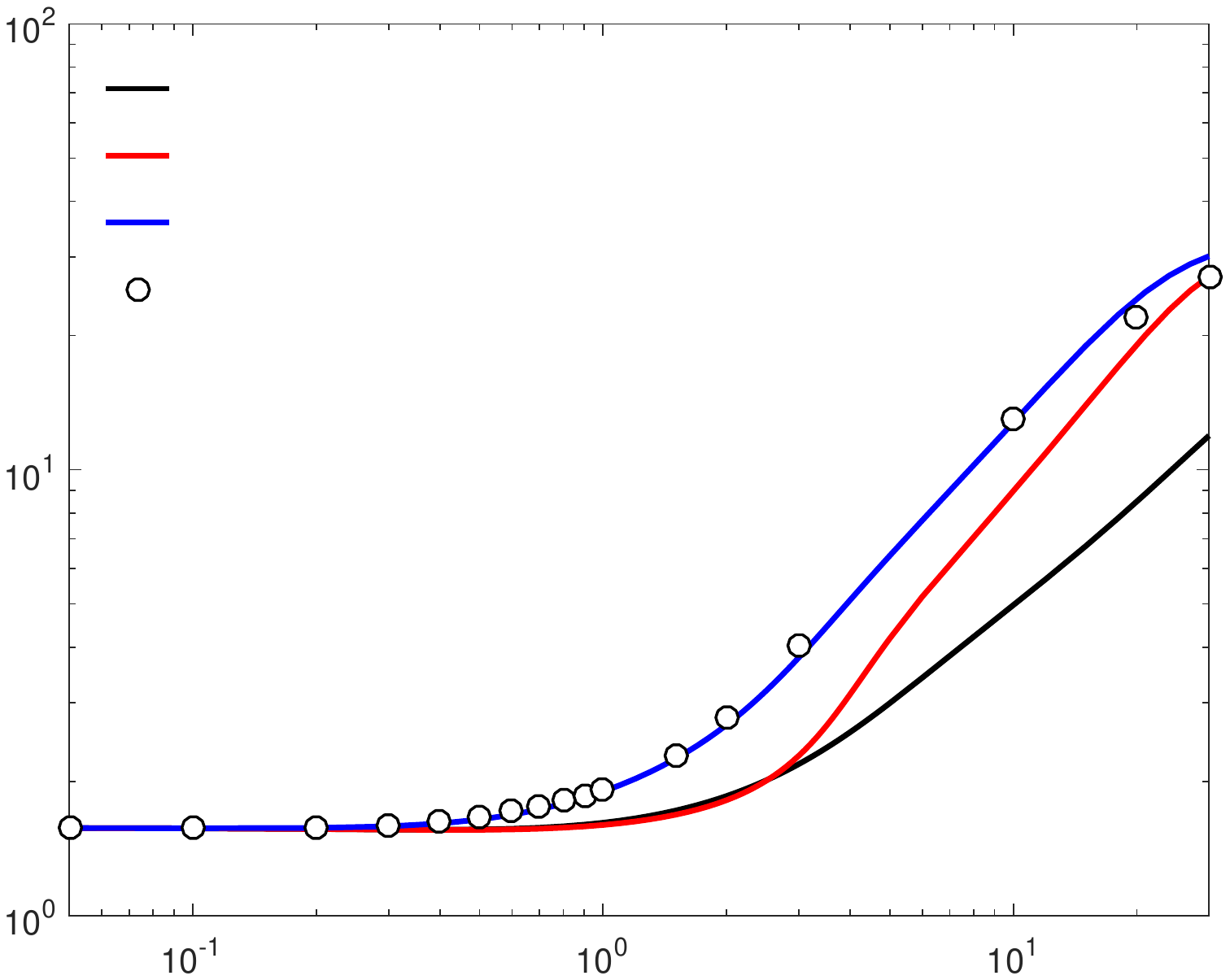}
\put(105,-2){$St$}
\put(0,75){\rotatebox{90}{$S^p_{2\parallel}/u_\eta^2$}}
\put(45,149){\small{PPM}}
\put(45,139){\small{PPM$^*$}}
\put(45,129){\small{PPM$^{**}$}}
\put(45,119){\small{DNS}}
\end{overpic}}
\subfloat[]
{\begin{overpic}
[trim = 20mm 70mm 24mm 60mm,scale=0.45,clip,tics=20]{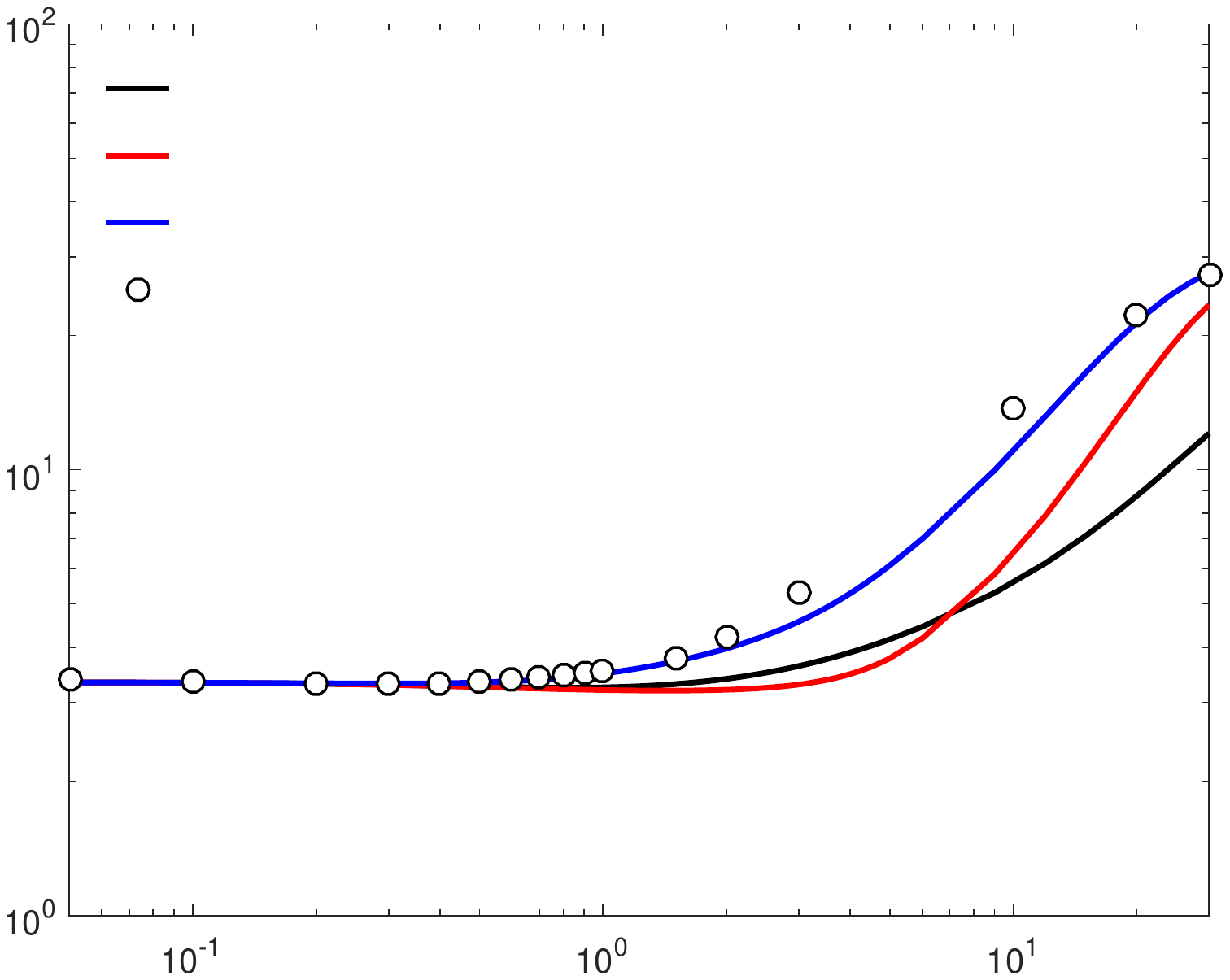}
\put(105,-2){$St$}
\put(0,75){\rotatebox{90}{$S^p_{2\parallel}/u_\eta^2$}}
\put(45,149){\small{PPM}}
\put(45,139){\small{PPM$^*$}}
\put(45,129){\small{PPM$^{**}$}}
\put(45,119){\small{DNS}}
\end{overpic}}
\caption{Comparison of the predictions from PPM, PPM$^{*}$ and PPM$^{**}$ with DNS data for $S^p_{2\parallel}$ as a function of $St$ and for (a) $r/\eta=0.15$, (b) $r/\eta=0.45$, (c) $r/\eta=1$, and (d) $r/\eta=3.25$, (e) $r/\eta=5.25$, (f) $r/\eta=8.25$.}
\label{Sp2plotb}
\end{figure}

We begin by first considering the results in Fig.~\ref{Sp2plotb} for PPM$^{*}$, which show that incorporating the correct BIT mean-square separation behavior for $s\in[0,s_c)$ leads to significant enhancement in the predicted values of $S^p_{2\parallel}$, when $St\gtrsim\mathcal{O}(1)$. This enhancement leads to improved predictions, such that PPM$^*$ compares more favorably with the DNS than PPM, as anticipated in \cite{bragg14c}. 

However, for each of the separations, there is a range of $St$ for which both PPM$^*$ and PPM systematically underpredict the DNS. The results in Fig.~\ref{Sp2plotb} reveal that PPM$^{**}$ performs significantly better than either PPM$^{*}$ or PPM when compared with the DNS, for $r\geq\eta$ and $St\lesssim 1$. For $St\lesssim 1$ and  $r<\eta/2$, PPM$^{**}$ overpredicts the DNS data. This is most likely because (\ref{r2larges}) assumes that the particles are in the inertial range, yet for $St\lesssim 1$, if $r$ is sufficiently small then $\sqrt{\langle\|\bm{r}^p(-\mathcal{O}(s_c))\|^2\rangle_{\bm{r}}}<\mathcal{O}(10\eta)$, i.e. in the dissipation range. For $St>1$, PPM$^{**}$ overpredicts the DNS data for each $r$ tested. This is most likely due to the fact that in these cases, the assumption that $St_r(s)\ll1$, which was made in deriving (\ref{r2larges}), is not sufficiently satisfied. However, consistent with this explanation, the overpredictions for $St>1$ reduce with increasing $r$, and the discrepancies are small for $r\geq3\eta$.

In order to address these remaining deficiencies in PPM$^{**}$, we would require a theoretical prediction for $\langle\|\bm{r}^p(-s)\|^2\rangle_{\bm{r}}$ that is valid when $\sqrt{\langle\|\bm{r}^p(-\mathcal{O}(s_c))\|^2\rangle_{\bm{r}}}<\mathcal{O}(10\eta)$, and that is valid for $St_r(s>s_c)\geq\mathcal{O}(1)$. These present significant theoretical challenges, especially since there is no obvious small-parameter to use in these regimes. Indeed, it is difficult to see how such predictions could be constructed without also introducing another integral equation for $\langle\|\bm{r}^p(-s)\|^2\rangle_{\bm{r}}$, in addition to that describing $\bm{S}^p_2$ in the Pan \& Padoan model.

%%
%\begin{figure}[ht]
%\centering
%\vspace{-18mm}
%\subfloat[]
%{\begin{overpic}
%[trim = 20mm 70mm 24mm 60mm,scale=0.45,clip,tics=20]{wr2_reta_015.pdf}
%\put(105,-2){$St$}
%\put(0,75){\rotatebox{90}{$S^p_{2\parallel}/u_\eta^2$}}
%\put(45,149){\small{PPM}}
%\put(45,139){\small{PPM$^*$}}
%\put(45,129){\small{DNS}}
%\end{overpic}}
%\subfloat[]
%{\begin{overpic}
%[trim = 20mm 70mm 24mm 60mm,scale=0.45,clip,tics=20]{wr2_reta_045.pdf}
%\put(105,-2){$St$}
%\put(0,75){\rotatebox{90}{$S^p_{2\parallel}/u_\eta^2$}}
%\put(45,149){\small{PPM}}
%\put(45,139){\small{PPM$^*$}}
%\put(45,129){\small{DNS}}
%\end{overpic}}\\
%\vspace{-10mm}
%\subfloat[]
%{\begin{overpic}
%[trim = 20mm 70mm 24mm 60mm,scale=0.45,clip,tics=20]{wr2_reta_1.pdf}
%\put(105,-2){$St$}
%\put(0,75){\rotatebox{90}{$S^p_{2\parallel}/u_\eta^2$}}
%\put(45,149){\small{PPM}}
%\put(45,139){\small{PPM$^*$}}
%\put(45,129){\small{DNS}}
%\end{overpic}}
%\subfloat[]
%{\begin{overpic}
%[trim = 20mm 70mm 24mm 60mm,scale=0.45,clip,tics=20]{wr2_reta_305.pdf}
%\put(105,-2){$St$}
%\put(0,75){\rotatebox{90}{$S^p_{2\parallel}/u_\eta^2$}}
%\put(45,149){\small{PPM}}
%\put(45,139){\small{PPM$^*$}}
%\put(45,129){\small{DNS}}
%\end{overpic}}
%\caption{Comparison of the predictions from PPM and PPM$^*$ with DNS data for $S^p_{2\parallel}$ as a function of $St$ and for (a) $r/\eta=0.15$, (b) $r/\eta=0.45$, (c) $r/\eta=1$, and (d) $r/\eta=3.15$.}
%\label{Sp2plot}
%\end{figure}
%\FloatBarrier
%%
%
\FloatBarrier
Up to this point, we have attributed the errors at $St\lesssim 1$ and  $r<\eta/2$ to the behavior of the closure model for $\langle\|\bm{r}^p(-s)\|^2\rangle_{\bm{r}}$ in the regime $s\in[s_c,s_d)$. There is, however, another possible explanation pertaining to the behavior of (\ref{BITclos}). In particular, (\ref{BITclos}) does not account for the effect of preferential sampling on the BIT dispersion of the particles. As explained in \cite{bragg16}, in deriving (\ref{BITclos}) we used the approximation
\begin{align}
\bm{S}^{fp}_2(\bm{r})\equiv\Big\langle\|\Delta\bm{u}(\bm{r}^p(0),0)\|^2\Big\rangle_{\bm{r}}\approx\bm{S}^f_2(\bm{r}),\label{PSapx}	
\end{align}
which amounts to ignoring the effects of preferential sampling (the approximation in (\ref{PSapx}) is in fact exact in the limits $St\to0$ and $St\to\infty$). When preferential sampling occurs, $\bm{S}^{fp}_2<\bm{S}^f_2$, and this could affect the BIT dispersion behavior of the inertial particles. 

In considering the effect of the approximation $\bm{S}^{fp}_2\approx \bm{S}^f_2$, we first note that the effects of preferential sampling on the BIT dispersion should become weaker as $r$ reduced. This may be demonstrated by noting that since in the dissipation range $\mathrm{tr}[\bm{S}^{p}_2]\propto r^{\zeta}$ \cite{gustavsson11,bragg14c}, then
\begin{align}
	\mathrm{tr}[\bm{S}^{p}_2]\Big/\mathrm{tr}[\bm{S}^{fp}_2]\propto r^{\zeta-2},
\end{align}
and since $\zeta(St>0)\in[0,2)$ \cite{ireland16a}, we then find  
\begin{align}
\begin{split}
\lim_{r\to0}\Big\langle\|\bm{r}^p(-s)\|^2\Big\rangle_{\bm{r}}\to r^2+\mathcal{G}^2(-s)\mathrm{tr}[\bm{S}^p_2].
\end{split}
\end{align}
This shows that in the limit $r\to0$, (\ref{BITclos}) is unaffected by the approximation $\bm{S}^{fp}_2\approx \bm{S}^f_2$ made in its derivation.

In order to examine the effect of $\bm{S}^{fp}_2\approx \bm{S}^f_2$ on the BIT dispersion when $r=\mathcal{O}(\eta)$, we compared the results from (\ref{BITclos}) with the predictions from (\ref{BITclos}) when $\bm{S}^f_2$ is replaced with $\bm{S}^{fp}_2$. By simply ``playing'' with the input value for $\bm{S}^{fp}_2$, we found that for $St\leq\mathcal{O}(1)$, the results were almost identical unless $\mathrm{tr}[\bm{S}^{fp}_2]\ll\mathrm{tr}[\bm{S}^f_2]$. However, DNS measurements have shown that $\forall St: \mathrm{tr}[\bm{S}^{fp}_2]=\mathcal{O}(\mathrm{tr}[\bm{S}^f_2])$ \cite{ireland16a}. Therefore, any discrepancies between PPM$^*$, PPM$^{**}$ and the DNS data for $St\lesssim 1$ and  $r<\eta/2$, cannot be caused by the approximation $\bm{S}^{fp}_2\approx \bm{S}^f_2$ made in deriving (\ref{BITclos}). This supports our earlier conclusion that these errors arise from errors in the prescription of $\langle\|\bm{r}^p(-s)\|^2\rangle_{\bm{r}}$ for the regime $s\in[s_c,s_d)$ in PPM$^*$ and PPM$^{**}$.
\begin{figure}[ht]
\centering
\vspace{-7mm}
\subfloat[]
{\begin{overpic}
[trim = 20mm 70mm 24mm 60mm,scale=0.45,clip,tics=20]{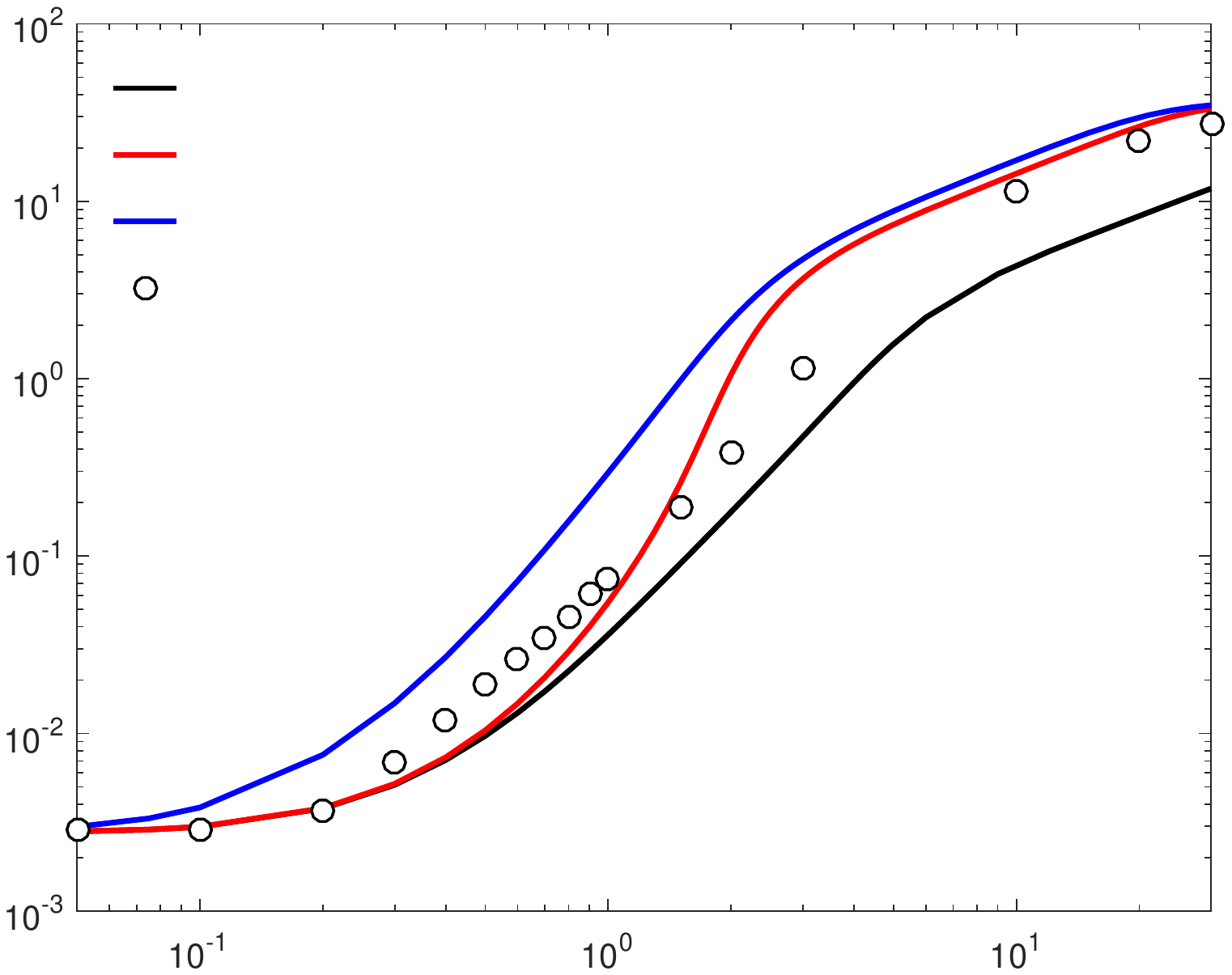}
\put(105,-2){$St$}
\put(0,75){\rotatebox{90}{$S^p_{2\perp}/u_\eta^2$}}
\put(45,149){\small{PPM}}
\put(45,139){\small{PPM$^*$}}
\put(45,129){\small{PPM$^{**}$}}
\put(45,119){\small{DNS}}
\end{overpic}}
\subfloat[]
{\begin{overpic}
[trim = 20mm 70mm 24mm 60mm,scale=0.45,clip,tics=20]{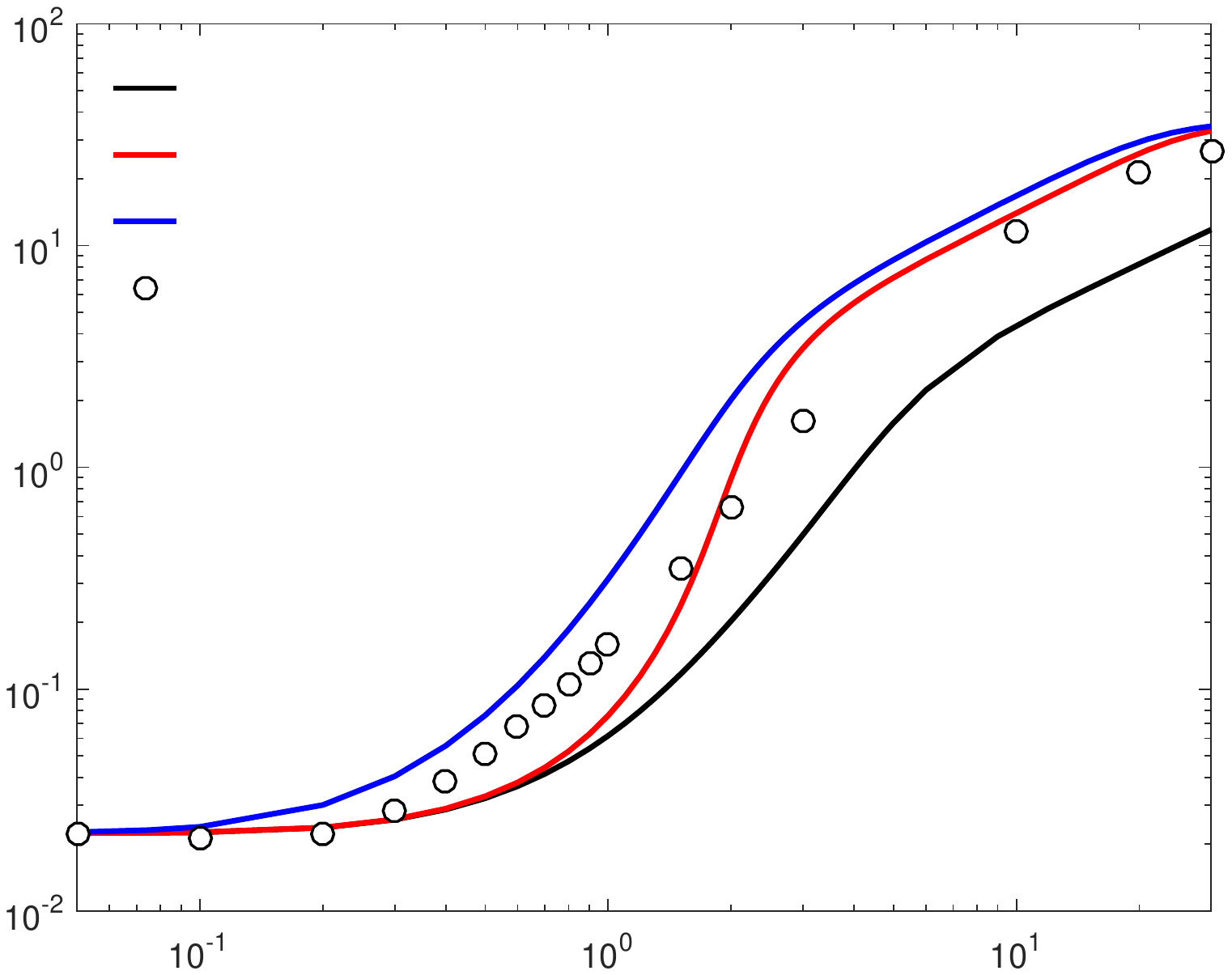}
\put(105,-2){$St$}
\put(0,75){\rotatebox{90}{$S^p_{2\perp}/u_\eta^2$}}
\put(45,149){\small{PPM}}
\put(45,139){\small{PPM$^*$}}
\put(45,129){\small{PPM$^{**}$}}
\put(45,119){\small{DNS}}
\end{overpic}}\\
\vspace{-10mm}
\subfloat[]
{\begin{overpic}
[trim = 20mm 70mm 24mm 60mm,scale=0.45,clip,tics=20]{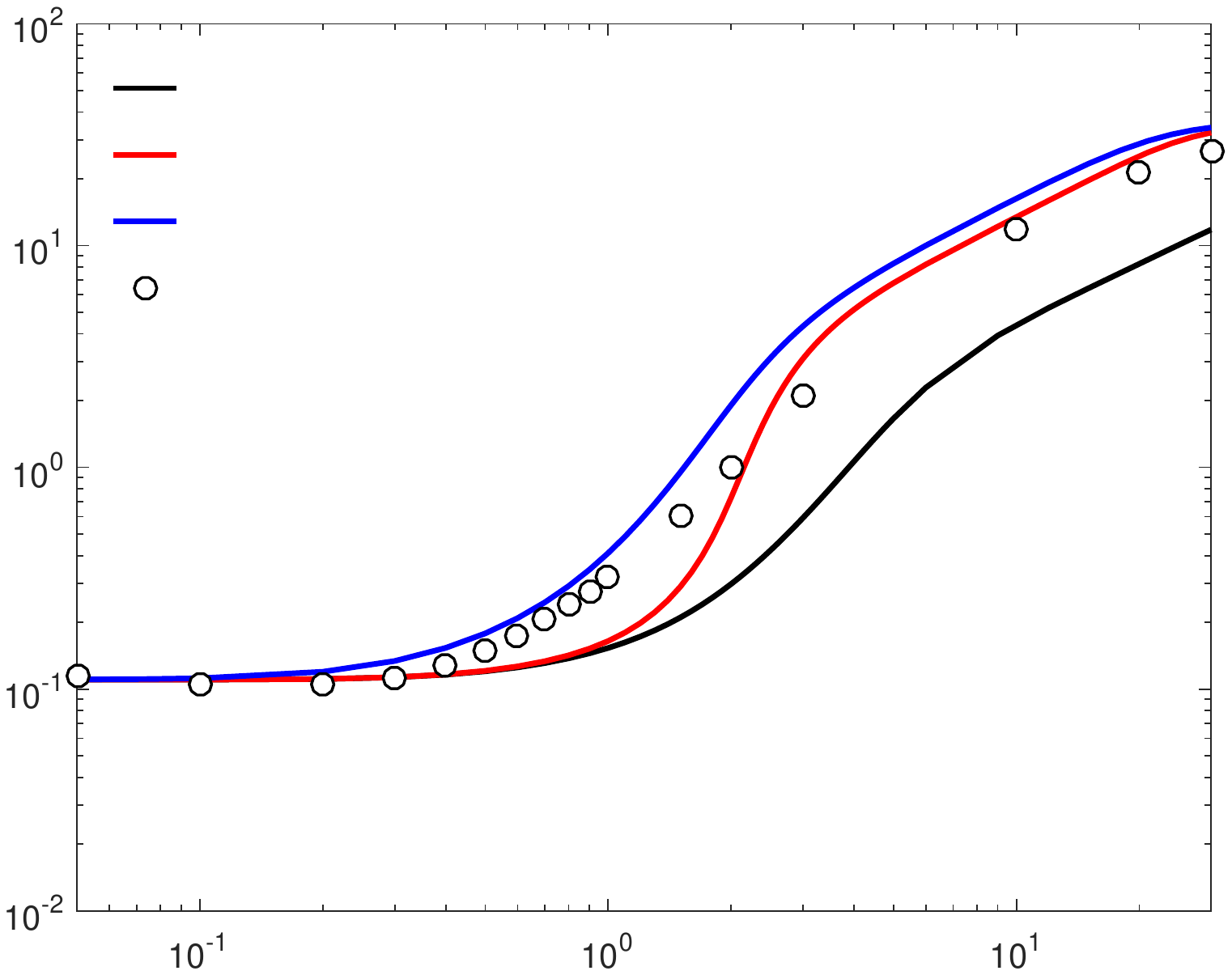}
\put(105,-2){$St$}
\put(0,75){\rotatebox{90}{$S^p_{2\perp}/u_\eta^2$}}
\put(45,149){\small{PPM}}
\put(45,139){\small{PPM$^*$}}
\put(45,129){\small{PPM$^{**}$}}
\put(45,119){\small{DNS}}
\end{overpic}}
\subfloat[]
{\begin{overpic}
[trim = 20mm 70mm 24mm 60mm,scale=0.45,clip,tics=20]{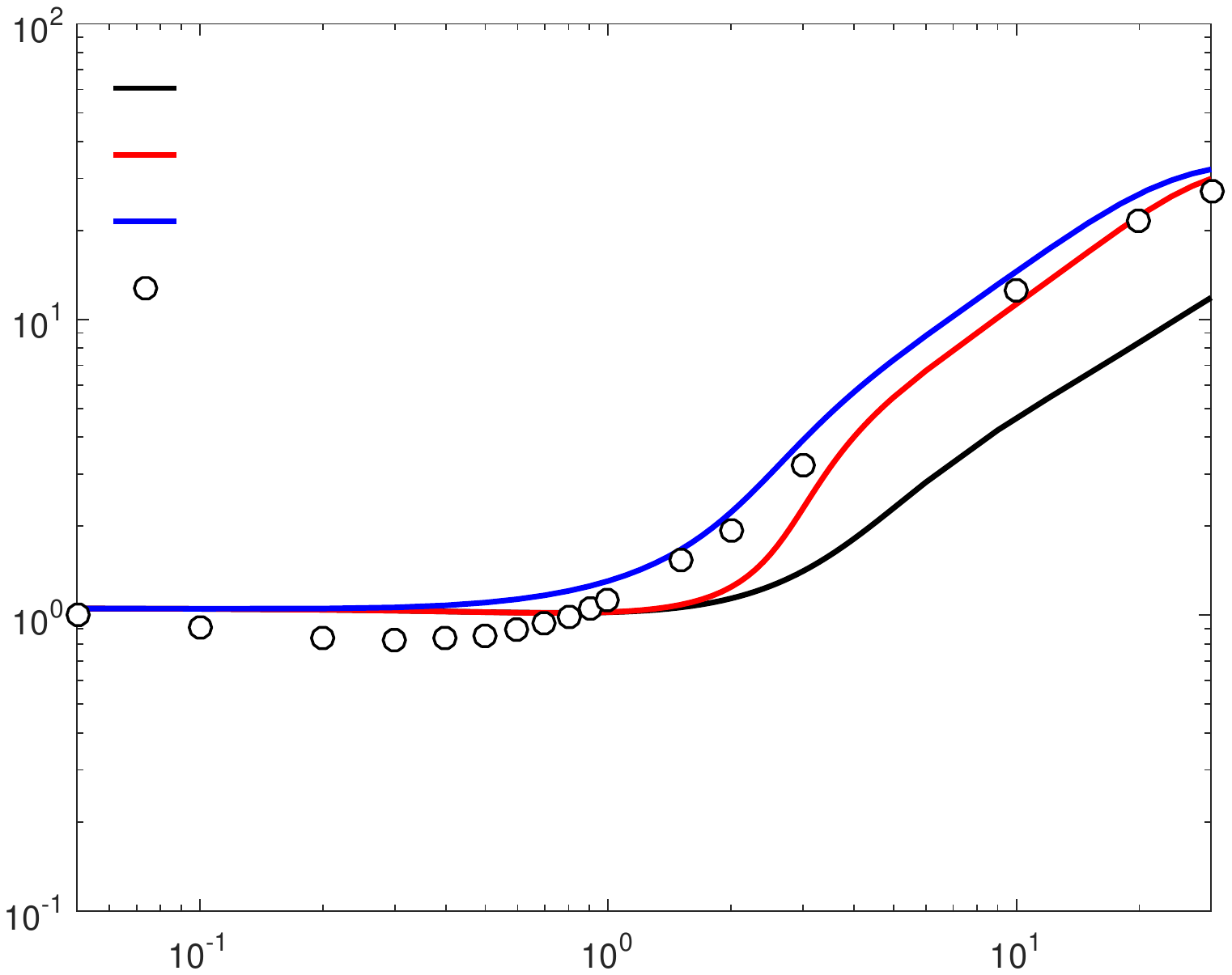}
\put(105,-2){$St$}
\put(0,75){\rotatebox{90}{$S^p_{2\perp}/u_\eta^2$}}
\put(45,149){\small{PPM}}
\put(45,139){\small{PPM$^*$}}
\put(45,129){\small{PPM$^{**}$}}
\put(45,119){\small{DNS}}
\end{overpic}}\\
\vspace{-10mm}
\subfloat[]
{\begin{overpic}
[trim = 20mm 70mm 24mm 60mm,scale=0.45,clip,tics=20]{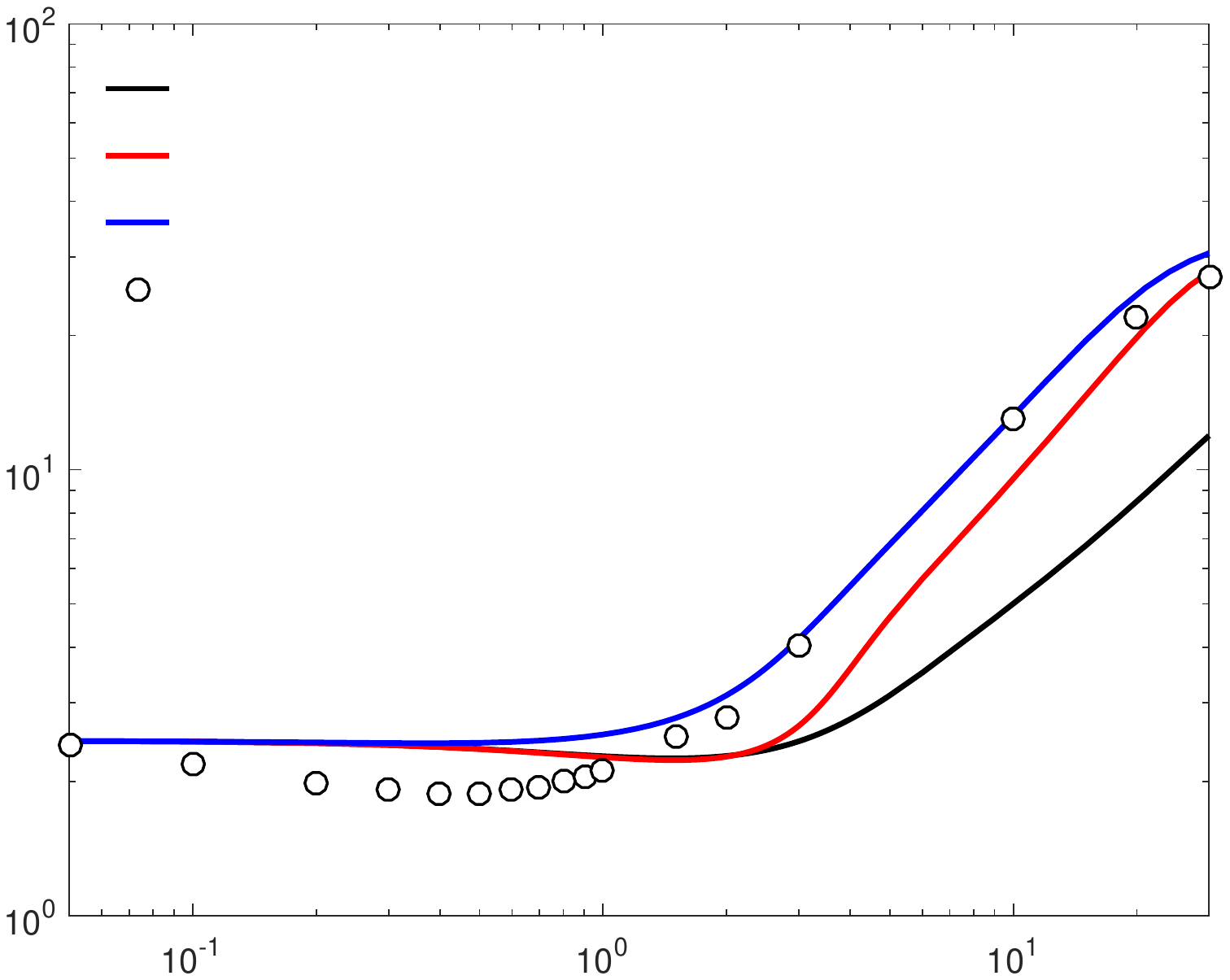}
\put(105,-2){$St$}
\put(0,75){\rotatebox{90}{$S^p_{2\perp}/u_\eta^2$}}
\put(45,149){\small{PPM}}
\put(45,139){\small{PPM$^*$}}
\put(45,129){\small{PPM$^{**}$}}
\put(45,119){\small{DNS}}
\end{overpic}}
\subfloat[]
{\begin{overpic}
[trim = 20mm 70mm 24mm 60mm,scale=0.45,clip,tics=20]{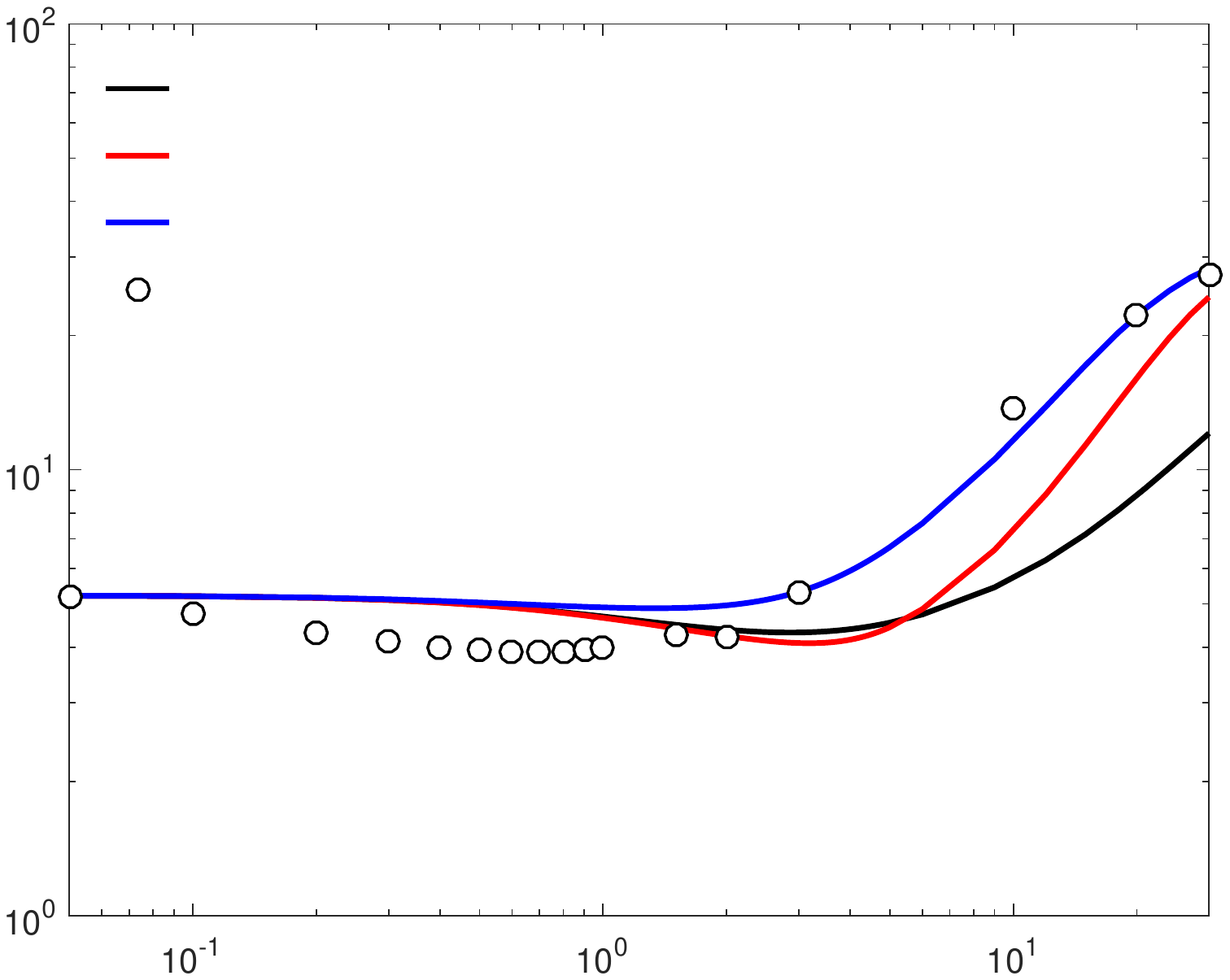}
\put(105,-2){$St$}
\put(0,75){\rotatebox{90}{$S^p_{2\perp}/u_\eta^2$}}
\put(45,149){\small{PPM}}
\put(45,139){\small{PPM$^*$}}
\put(45,129){\small{PPM$^{**}$}}
\put(45,119){\small{DNS}}
\end{overpic}}
\caption{Comparison of the predictions from PPM, PPM$^{*}$ and PPM$^{**}$ with DNS data for $S^p_{2\perp}$ as a function of $St$ and for (a) $r/\eta=0.15$, (b) $r/\eta=0.45$, (c) $r/\eta=1$, and (d) $r/\eta=3.25$, (e) $r/\eta=5.25$, (f) $r/\eta=8.25$.}
\label{Sp2transplotb}
\end{figure}
\FloatBarrier
When we turn to consider the results for $S^p_{2\perp}$ in Fig.~\ref{Sp2transplotb}, we find that the models perform about as well as they do in predicting $S^p_{2\parallel}$, with one important exception. The DNS results in Fig.~\ref{Sp2transplotb} show that $S^p_{2\perp}$ initially decreases with increasing $St$ when $r\gtrsim \eta$, and the models either fail to predict this, or else predict that the decrease begins at too large a value of $St$. As explained in \cite{ireland16a}, the initial decrease of $S^p_{2\perp}$ with increasing $St$ occurs because of preferential sampling, which affects $S^p_{2\perp}$ much more than $S^p_{2\parallel}$ since the particles undersample high-rotation-rate regions more than they undersample high-strain-rate regions. By undersampling regions with high-rotation-rate, the inertial particles experience, on average, perpendicular fluid relative velocities that are weaker than those experienced by fluid particles.

That PPM, PPM$^*$ and PPM$^{**}$ all fail to correctly predict this feature is not due to their approximations for $\langle\|\bm{r}^p(-s)\|^2\rangle_{\bm{r}}$, but due to an underlying assumption in the Pan \& Padoan modeling framework that neglects the effects of preferential sampling at the explicit level. Indeed, it is straightforward to show that the integral equation defining $\bm{S}^p_2$ in \cite{pan10} gives 
\begin{align}
\bm{S}^p_2(\bm{r})=\bm{S}^f_2+\mathcal{O}(St),\quad\text{for}\,St\ll1,
\end{align}
whereas the correct behavior is 
\begin{align}
\bm{S}^p_2(\bm{r})=\bm{S}^{fp}_2+\mathcal{O}(St),\quad\text{for}\,St\ll1.
\end{align}
Note that the results in Fig.~\ref{Sp2transplotb} show that the effect of the preferential sampling on $S^p_{2\perp}$ becomes more significant as $r$ is increased. This is because as $r$ is increased, the role of the path-history effects weakens, allowing the preferential sampling to play a more significant role in determining $S^p_{2\perp}$ \cite{bragg14c,ireland16a}.
\section{Implications for predicting the RDF}\label{rdfimp}
We now consider the implications of our current ability to predict $\bm{S}^{p}_2$ for predicting the RDF, $g(r)$, that quantifies the level of the spatial clustering of inertial particles in turbulence \cite{sundaram4}. 

In \cite{bragg14b,ireland16a} we showed that when DNS data for $\bm{S}^p_2$ is used in the transport equation for $g(r)$ from the Zaichik \& Alipchenkov Model (ZAM hereafter) \cite{zaichik09}, the RDF can be accurately predicted. The ZAM cannot itself accurately predict the RDF, except for $St\lesssim 0.3$ \cite{ireland16a}, since its predictions for $\bm{S}^p_2$ are in gross error when $St\geq\mathcal{O}(1)$ \cite{bragg14c}. In \cite{bragg14c} we suggested that, subject to improvements in its predictive capabilities, the Pan \& Padoan model might provide a promising alternative way to predict $\bm{S}^p_2$ when $St\geq\mathcal{O}(1)$, which when coupled with the ZAM transport equation for $g(r)$, could provide a way to predict the RDF. We now consider in more detail the ability of PPM, PPM$^*$ and PPM$^{**}$ to predict the RDF via this method.

For a statistically stationary, isotropic system, the ZAM transport equation for $g(r)$ is
\begin{align}
	0=-\tau_p\Big(\lambda_\parallel + S^p_{2\parallel}\Big)\nabla_r g-\tau_p g\Big(\nabla_r S^p_{2\parallel}+2r^{-1}\Big[S^p_{2\parallel}-S^p_{2\perp}\Big]\Big),\label{zameq}
\end{align}
where $\lambda_\parallel (r)$ is a diffusion coefficient that describes the non-Markovian effect of the local turbulence on the diffusion of the particles, and is given by
\begin{align}
\lambda_\parallel =\frac{\tau_r^2 S^f_{2\parallel}}{\tau_p(\tau_r+\tau_p)},
\end{align}
where $\tau_r$ is the eddy-turnover timescale at separation $r$ (see \cite{zaichik09} for the formula describing $\tau_r$). Equation (\ref{zameq}) is solved with the boundary condition $g(r\to\infty)\to1$.

In the dissipation range (in particular, $r\to0$), the RDF is known to have the scale-invariant form $g(r)\propto r^{-\xi}$, where $\xi(St)\in[0,1)$ \cite{bec07,ireland16a}. In order for the solution of (\ref{zameq}) to posses this form we must have
\begin{align}
\Big(\lambda_\parallel + S^p_{2\parallel}\Big)^{-1}\Big(\nabla_r S^p_{2\parallel}+2r^{-1}\Big[S^p_{2\parallel}-S^p_{2\perp}\Big]\Big)\propto r^{-1}.\label{SIreq}
\end{align}
In the dissipation range, $\lambda_\parallel \propto r^2$, and (\ref{SIreq}) is satisfied in the limit $r\to0$ provided that $S^p_{2\parallel}$ and $S^p_{2\perp}$ posses the scale-invariant forms $S^p_{2\parallel}\propto r^{\zeta_\parallel}$, $S^p_{2\perp}\propto r^{\zeta_\perp}$ with $\zeta_\parallel\in[0,2]$, $\zeta_\perp\in[0,2]$ and $\vert\zeta_\parallel-\zeta_\perp\vert\ll1$. These conditions, which are known to be satisfied by theoretical and numerical results \cite{gustavsson11,bragg14c,ireland16a}, place stringent requirements on the PPM (and its variants PPM$^{*}$ and PPM$^{**}$), if it is to be used in conjunction with (\ref{zameq}) to predict $g(r)$. Indeed, any model that does not predict scale-invariant forms for $S^p_{2\parallel}$ and $S^p_{2\perp}$ in the dissipation range will necessarily lead to predictions for $g(r)$, through (\ref{zameq}), that are both qualitatively and quantitatively wrong. 

For $St\ll1$, the PPM is guaranteed to generate scale-invariant predictions for $S^p_{2\parallel}$ and $S^p_{2\perp}$ in the dissipation range, since $S^f_{2\parallel}\propto S^f_{2\perp}\propto r^2$. However, for $St\geq\mathcal{O}(1)$, the $r$ dependence of $S^p_{2\parallel}$ and $S^p_{2\perp}$ predicted by PPM depends in a very complicated way upon time integrals of Lagrangian functions appearing in the integrand defining the model. These Lagrangian functions involve a number of closure approximations, and as a result, scale-invariance of the predicted forms of $S^p_{2\parallel}$ and $S^p_{2\perp}$ is not guaranteed.
\begin{figure}[ht]
\centering
\vspace{-7mm}
\subfloat[]
{\begin{overpic}
[trim = 20mm 70mm 15mm 60mm,scale=0.45,clip,tics=20]{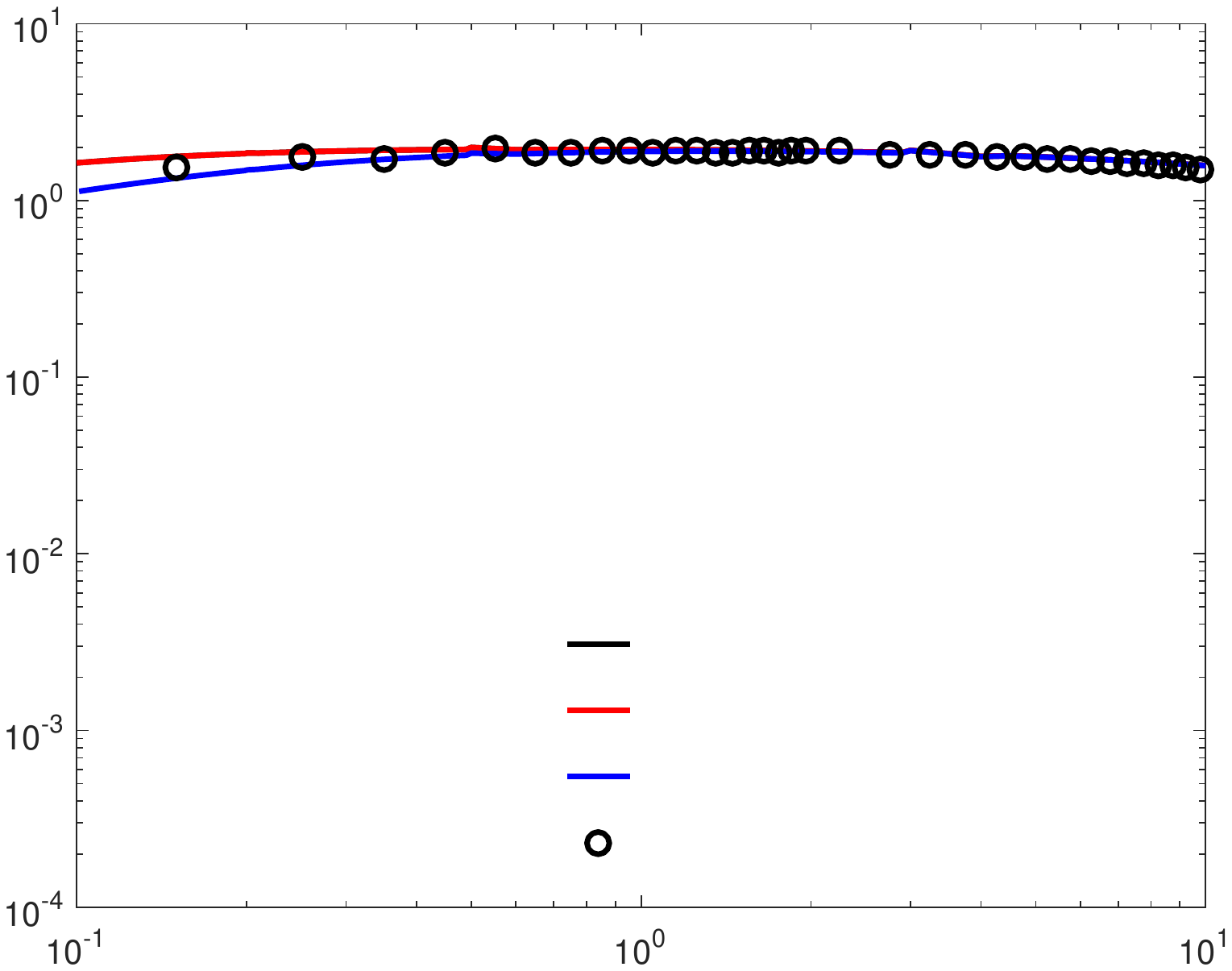}
\put(107,0){$r/\eta$}
\put(0,83){\rotatebox{90}{$\mathcal{M}_\parallel$}}
\put(120,60){\small{PPM}}
\put(120,49){\small{PPM$^*$}}
\put(120,38){\small{PPM$^{**}$}}
\put(120,27){\small{DNS}}
\end{overpic}}
\subfloat[]
{\begin{overpic}
[trim = 20mm 70mm 15mm 60mm,scale=0.45,clip,tics=20]{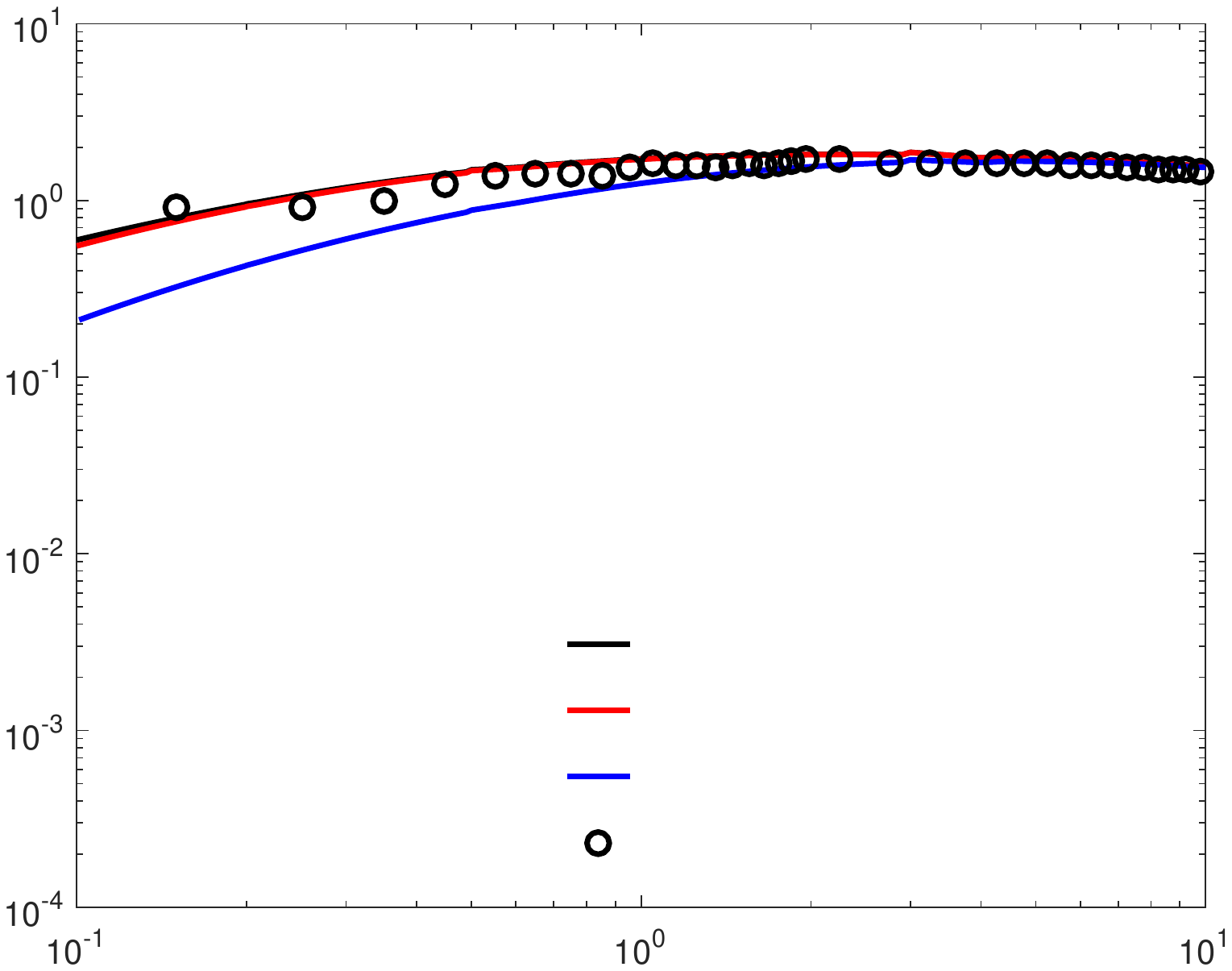}
\put(107,0){$r/\eta$}
\put(0,83){\rotatebox{90}{$\mathcal{M}_\parallel$}}
\put(120,60){\small{PPM}}
\put(120,49){\small{PPM$^*$}}
\put(120,38){\small{PPM$^{**}$}}
\put(120,27){\small{DNS}}
\end{overpic}}\\
\vspace{-10mm}
\subfloat[]
{\begin{overpic}
[trim = 20mm 70mm 15mm 60mm,scale=0.45,clip,tics=20]{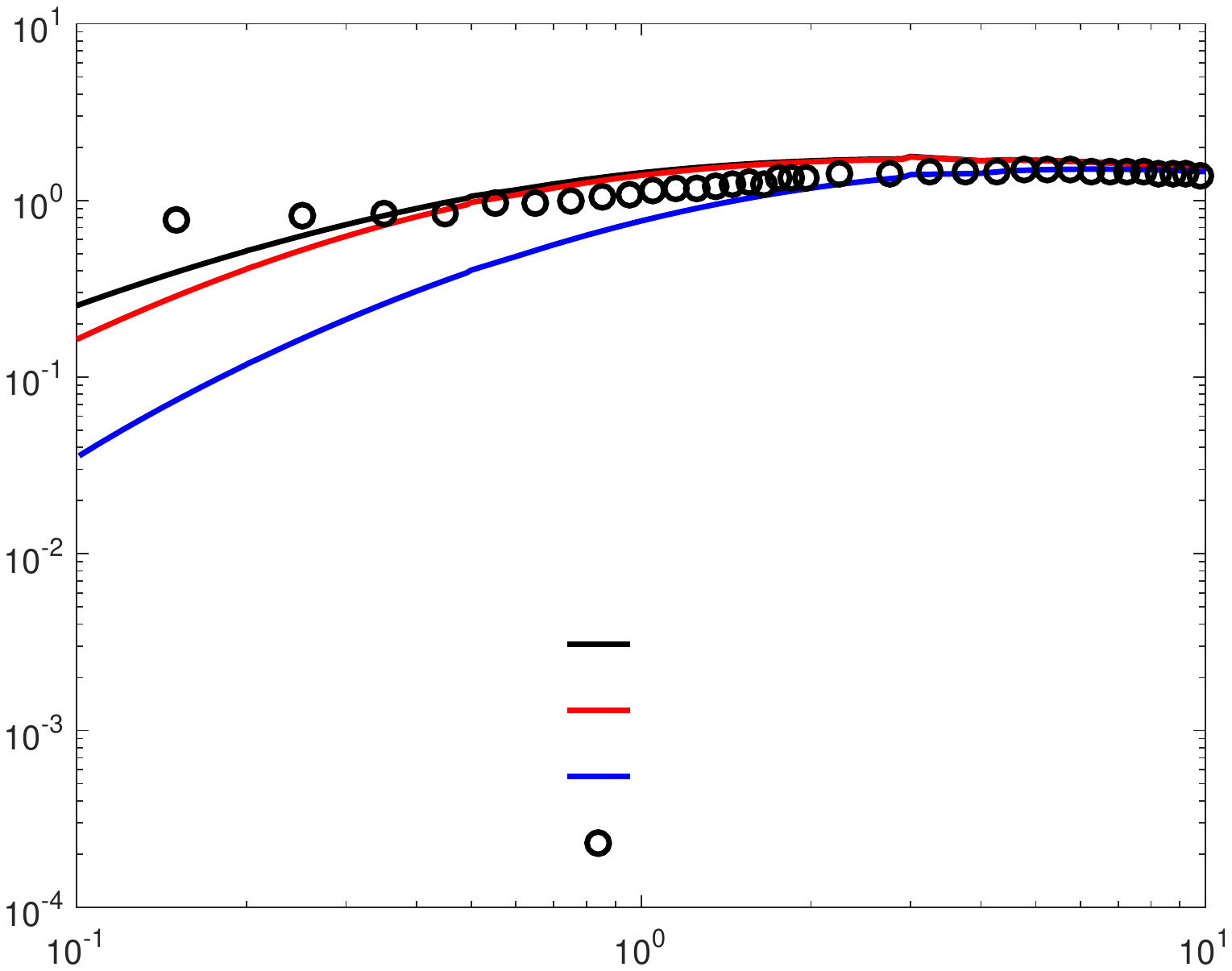}
\put(107,0){$r/\eta$}
\put(0,83){\rotatebox{90}{$\mathcal{M}_\parallel$}}
\put(120,60){\small{PPM}}
\put(120,49){\small{PPM$^*$}}
\put(120,38){\small{PPM$^{**}$}}
\put(120,27){\small{DNS}}
\end{overpic}}
\subfloat[]
{\begin{overpic}
[trim = 20mm 70mm 15mm 60mm,scale=0.45,clip,tics=20]{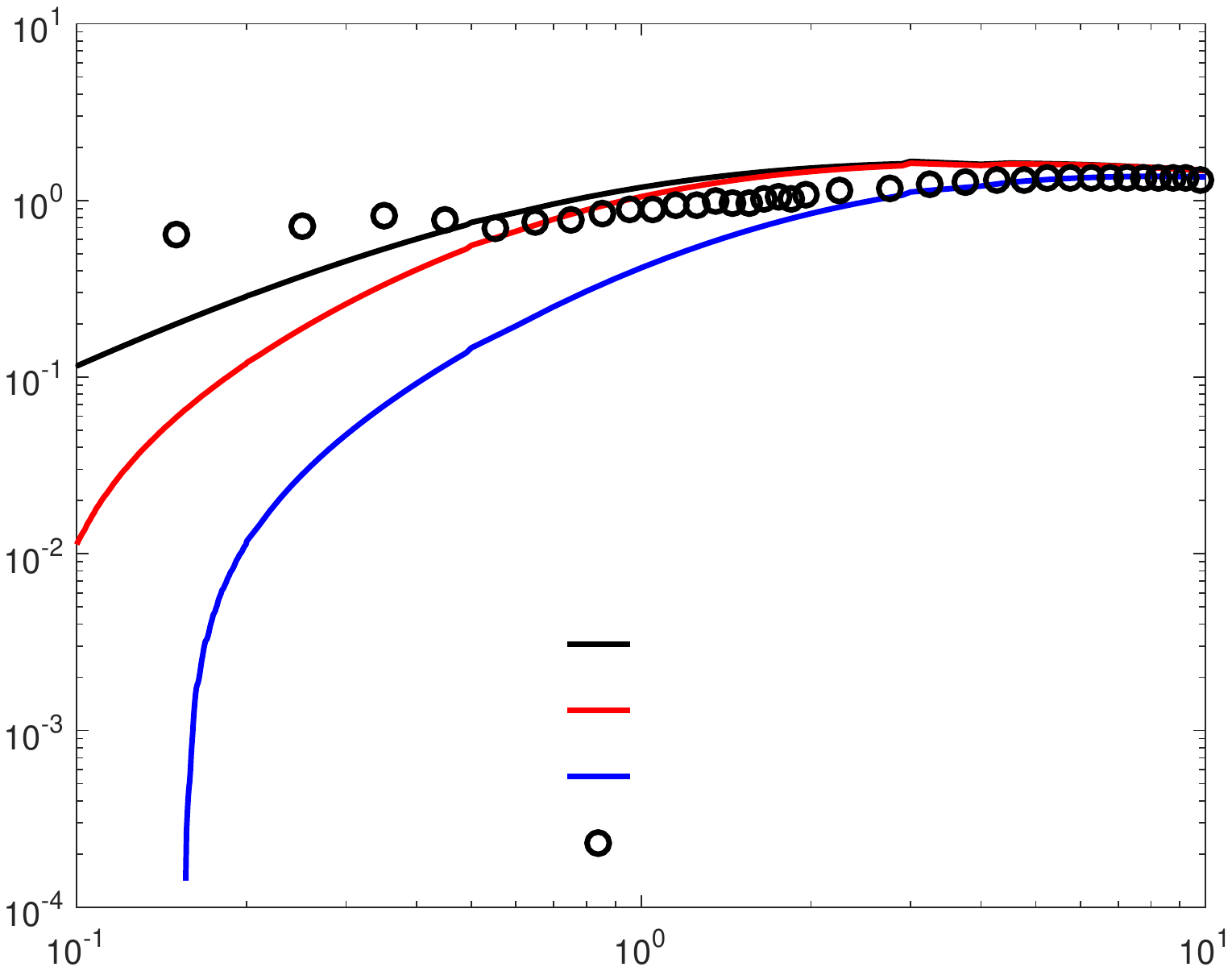}
\put(107,0){$r/\eta$}
\put(0,83){\rotatebox{90}{$\mathcal{M}_\parallel$}}
\put(120,60){\small{PPM}}
\put(120,49){\small{PPM$^*$}}
\put(120,38){\small{PPM$^{**}$}}
\put(120,27){\small{DNS}}
\end{overpic}}
\caption{Comparison of the predictions from PPM, PPM$^{*}$ and PPM$^{**}$ with DNS data for $\mathcal{M}_\parallel$ as a function of $r$ and for (a) $St=0.1$, (b) $St=0.4$, (c) $St=0.7$, and (d) $St=1$.}
\label{M_ll}
\end{figure}
\FloatBarrier
In order to examine this further, we consider the quantities
\begin{align}
\mathcal{M}_\parallel(r)&\equiv \frac{r \nabla_r S^p_{2\parallel}}{S^p_{2\parallel}},\\
\mathcal{M}_\perp(r)&\equiv \frac{r \nabla_r S^p_{2\perp}}{S^p_{2\perp}}.		
\end{align}
In the regime where $S^p_{2\parallel}$ and $S^p_{2\perp}$ are scale-invariant, $\mathcal{M}_\parallel$ and $\mathcal{M}_\perp$ are constants and take on values ${\mathcal{M}_\parallel\in[\xi,2]}$, ${\mathcal{M}_\perp\in[\xi,2]}$. Therefore, deviations of $\nabla_r\mathcal{M}_\parallel$ and $\nabla_r\mathcal{M}_\perp$ from zero provide a measure of the degree to which the models  fail to predict scale-invariant forms for $S^p_{2\parallel}$ and $S^p_{2\perp}$.
\begin{figure}[ht]
\centering
\vspace{-7mm}
\subfloat[]
{\begin{overpic}
[trim = 20mm 70mm 15mm 60mm,scale=0.45,clip,tics=20]{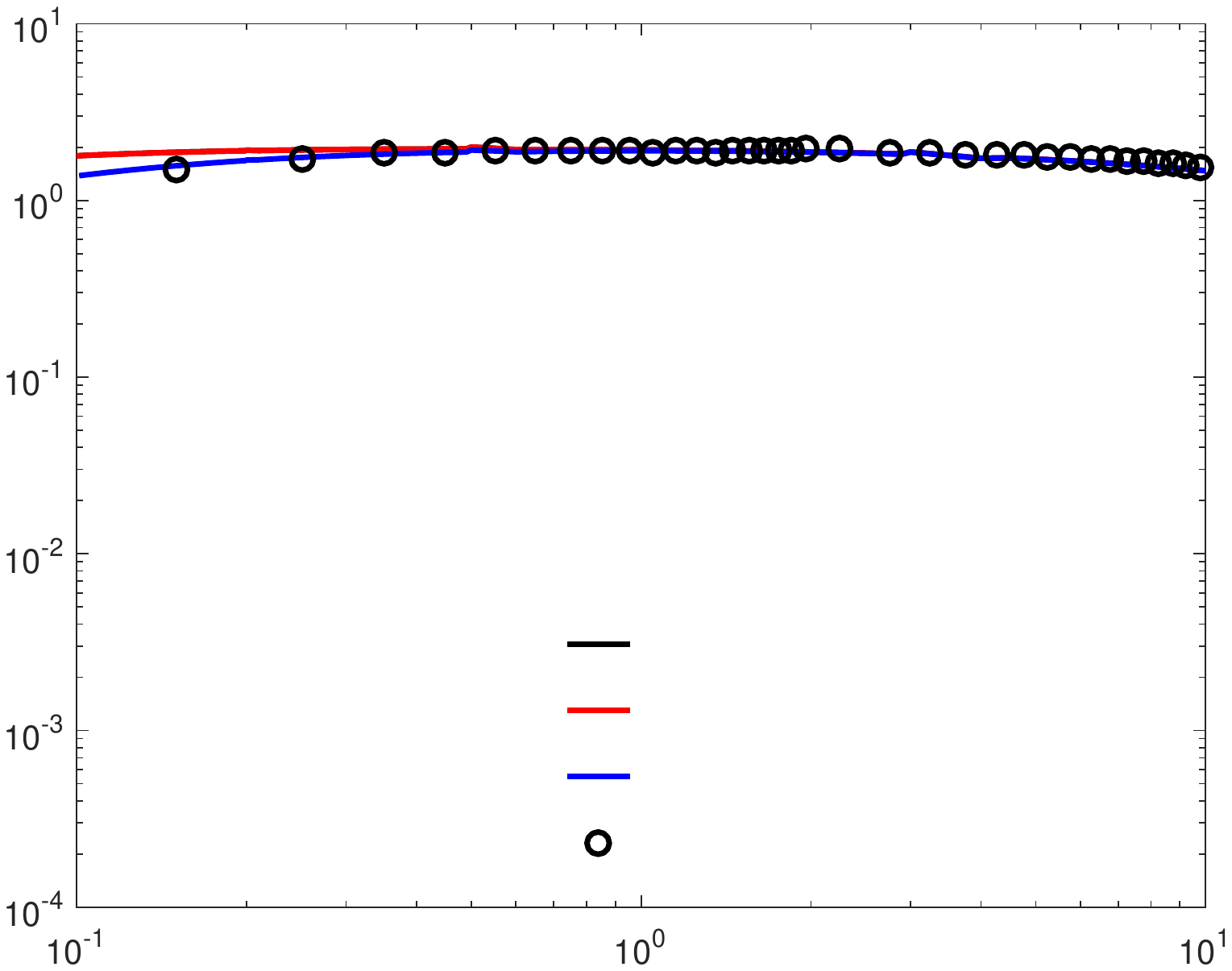}
\put(107,0){$r/\eta$}
\put(0,83){\rotatebox{90}{$\mathcal{M}_\perp$}}
\put(120,60){\small{PPM}}
\put(120,49){\small{PPM$^*$}}
\put(120,38){\small{PPM$^{**}$}}
\put(120,27){\small{DNS}}
\end{overpic}}
\subfloat[]
{\begin{overpic}
[trim = 20mm 70mm 15mm 60mm,scale=0.45,clip,tics=20]{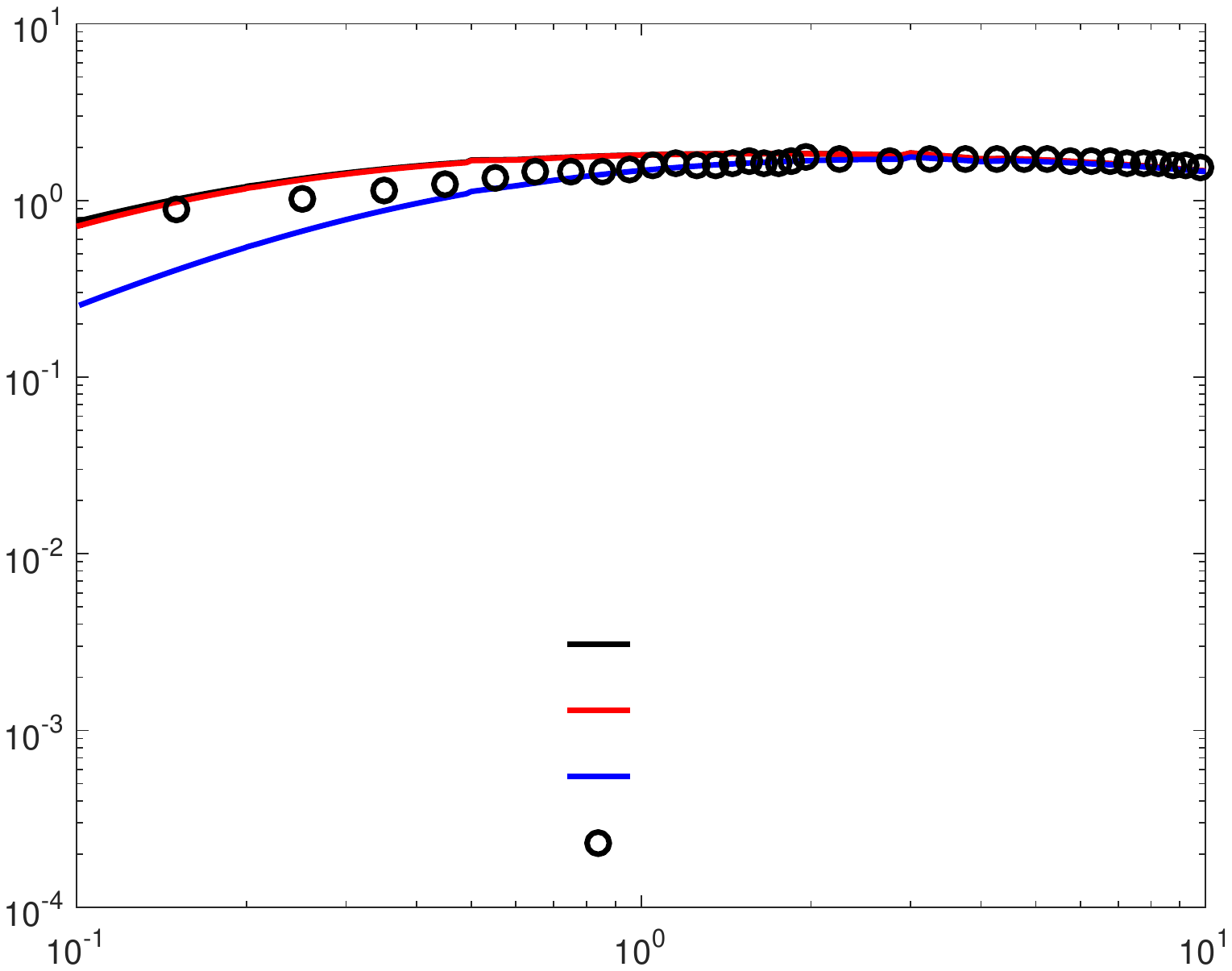}
\put(107,0){$r/\eta$}
\put(0,83){\rotatebox{90}{$\mathcal{M}_\perp$}}
\put(120,60){\small{PPM}}
\put(120,49){\small{PPM$^*$}}
\put(120,38){\small{PPM$^{**}$}}
\put(120,27){\small{DNS}}
\end{overpic}}\\
\vspace{-10mm}
\subfloat[]
{\begin{overpic}
[trim = 20mm 70mm 15mm 60mm,scale=0.45,clip,tics=20]{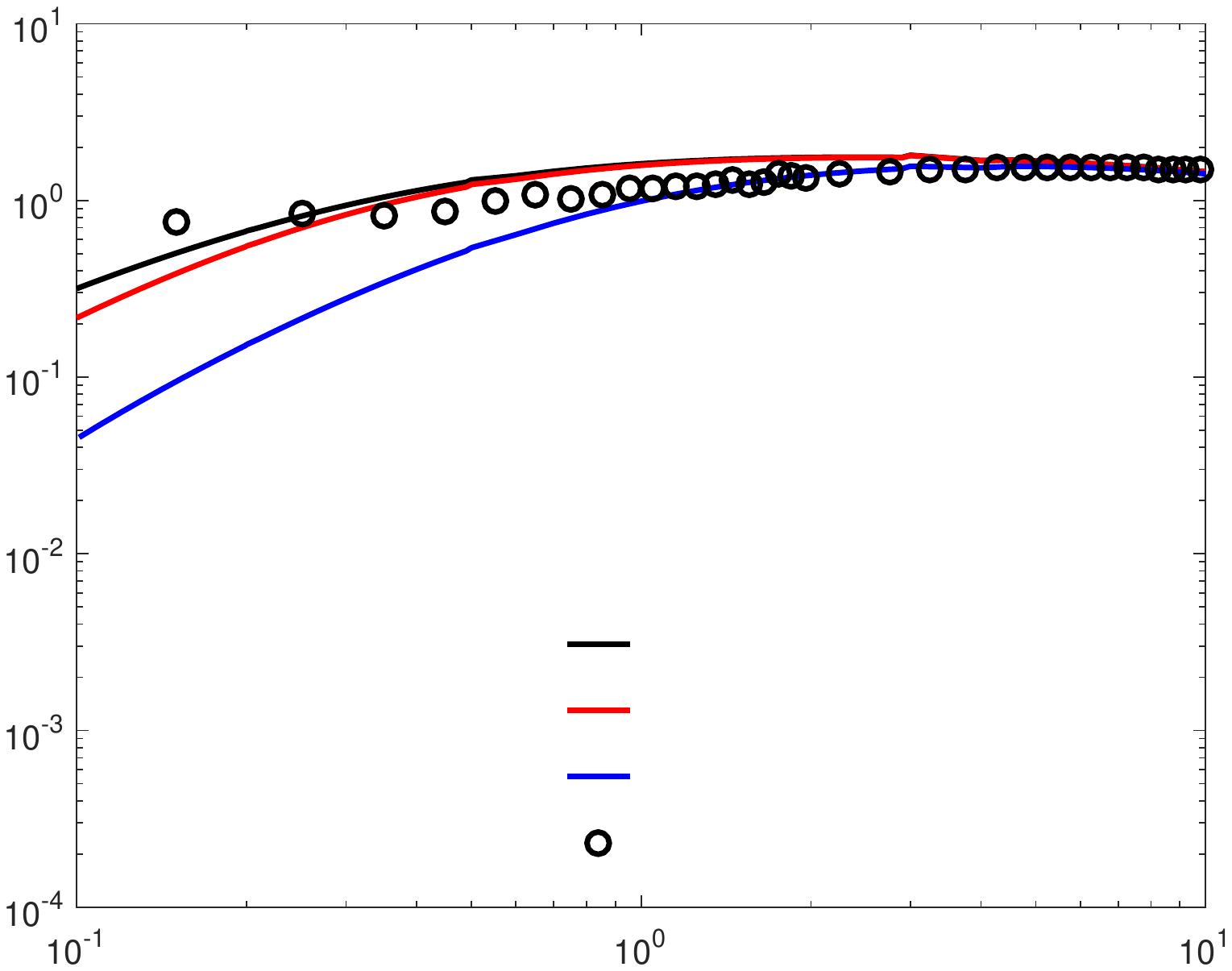}
\put(107,0){$r/\eta$}
\put(0,83){\rotatebox{90}{$\mathcal{M}_\perp$}}
\put(120,60){\small{PPM}}
\put(120,49){\small{PPM$^*$}}
\put(120,38){\small{PPM$^{**}$}}
\put(120,27){\small{DNS}}
\end{overpic}}
\subfloat[]
{\begin{overpic}
[trim = 20mm 70mm 15mm 60mm,scale=0.45,clip,tics=20]{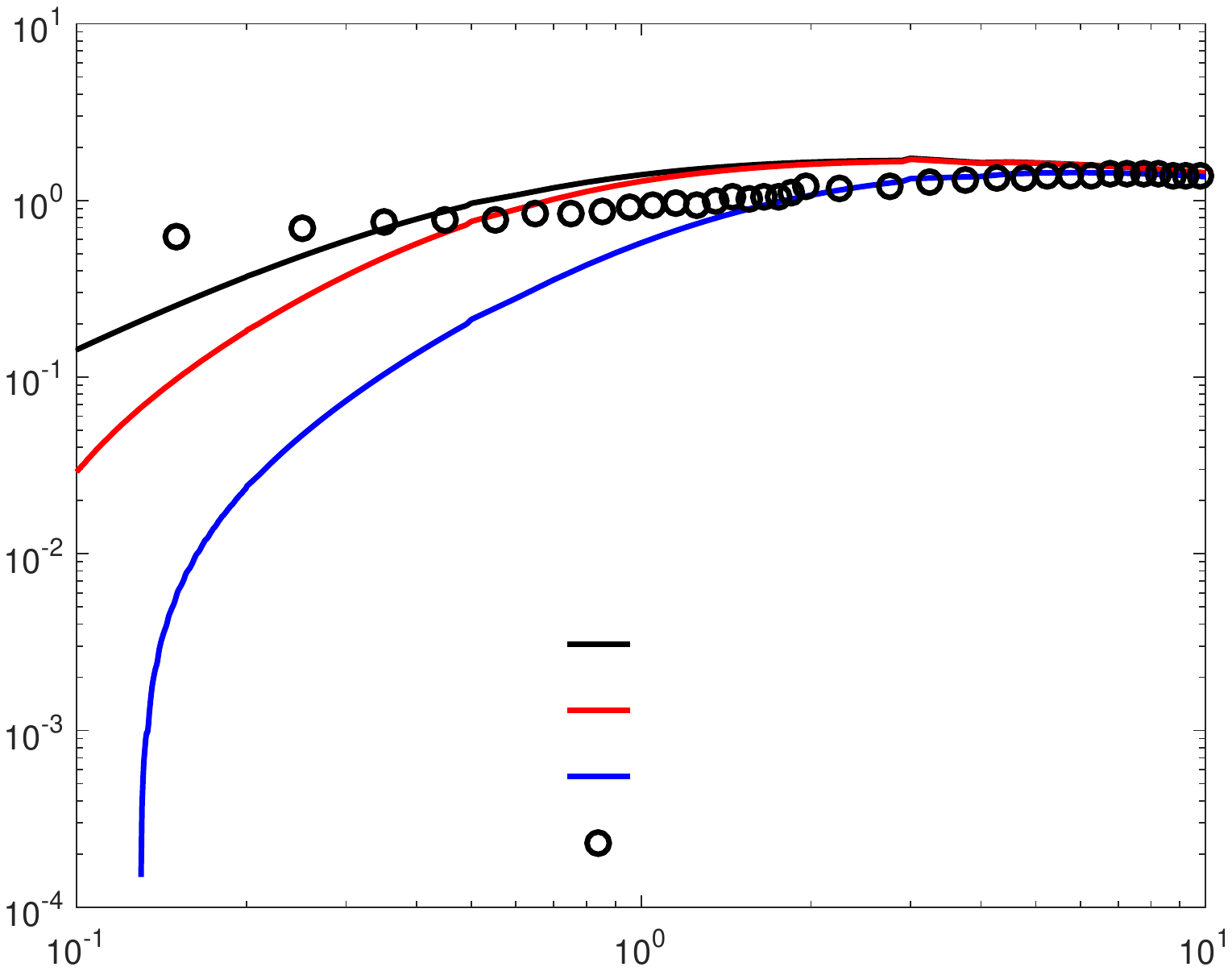}
\put(107,0){$r/\eta$}
\put(0,83){\rotatebox{90}{$\mathcal{M}_\perp$}}
\put(120,60){\small{PPM}}
\put(120,49){\small{PPM$^*$}}
\put(120,38){\small{PPM$^{**}$}}
\put(120,27){\small{DNS}}
\end{overpic}}
\caption{Comparison of the predictions from PPM, PPM$^{*}$ and PPM$^{**}$ with DNS data for $\mathcal{M}_\perp$ as a function of $r$ and for (a) $St=0.1$, (b) $St=0.4$, (c) $St=0.7$, and (d) $St=1$.}
\label{M_nn}
\end{figure}
\FloatBarrier
In Fig.~\ref{M_ll} and Fig.~\ref{M_nn} we compare the predictions for $\mathcal{M}_\parallel$ and $\mathcal{M}_\perp$ from PPM, PPM$^{*}$ and PPM$^{**}$ with DNS data. The results show that whereas the DNS data gives $\nabla_r\mathcal{M}_\parallel\approx 0$ and $\nabla_r\mathcal{M}_\perp\approx 0$ for $r\leq\mathcal{O}(\eta)$, the predictions from PPM, PPM$^{*}$ and PPM$^{**}$ do not (except for PPM, PPM$^{*}$ at $St=0.1$). The deviations of the model predictions from $\nabla_r\mathcal{M}_\parallel= 0$ and $\nabla_r\mathcal{M}_\perp= 0$  are very strong for $St\geq\mathcal{O}(1)$, and show that the model predictions for $S^p_{2\parallel}$ and $S^p_{2\perp}$ are very far from being scale-invariant. The results also show that for $r<\eta$, the predictions from PPM$^{**}$ for $\mathcal{M}_\parallel$ and $\mathcal{M}_\perp$ are in much poorer agreement with the DNS than PPM. This is related to the earlier observation that the results in Fig.~\ref{Sp2plotb} show that PPM$^{**}$ is accurate for $St\leq\mathcal{O}(1)$ when $r\geq\eta$, but leads to overpredictions when $r<\eta$. The consequence of this is that $\nabla_r S^p_{2\parallel}$, predicted by PPM$^{**}$, is too small for $r\leq\eta$.

These results therefore show that neither PPM nor its variants, are sufficiently accurate to be used in conjunction with (\ref{zameq}) to predict $g(r)$ for $St=\mathcal{O}(1)$. In particular, PPM, PPM$^{*}$ and PPM$^{**}$ are in gross qualitative error for $r\leq\mathcal{O}(\eta)$ and $St=\mathcal{O}(1)$.

It remains to be seen whether addressing the deficiencies in the closure model for $\langle\|\bm{r}^p(-s)\|^2\rangle_{\bm{r}}$ in PPM$^{**}$, discussed in \S\ref{Sp2results}, will be sufficient to generate predictions for $S^p_{2\parallel}$ and $S^p_{2\perp}$ that are scale-invariant (or at least sufficiently close to being so). Such challenges are left to future work.

\section{Conclusions}\label{conc}
In this paper we have used our the recently developed backward-in-time (BIT) relative dispersion theory for inertial particles in turbulence \cite{bragg16} to develop the theoretical model by Pan \& Padoan \cite{pan10} for predicting the relative velocities of inertial particles in turbulence. By comparing the model predictions with DNS data, we have shown that incorporating the BIT dispersion theory into the Pan \& Padoan model can lead to significant improvements compared to the original model in \cite{pan10}, showing good agreement with the DNS data for $St>\mathcal{O}(1)$. For the parallel component of the relative velocities, there is excellent agreement between the modified model and the DNS for $St\leq 1$ when $r\geq\eta$. The sources of error for $St\leq 1$ and $r<\eta$ are connected to limitations in the BIT dispersion theory, highlighting specific problems that need to be solved in future work. For the perpendicular component of the relative velocities, the models are inaccurate for $St\leq 1$ and $r\geq\eta$, and we argued that this is because the Pan \& Padoan modeling framework ignores the effect of preferential sampling.

We then considered how the Pan \& Padoan model could be used in conjunction with the transport equation for the RDF derived in \cite{zaichik09}. Predicting the RDF in this way places very specific constraints on the accuracy of the Pan \& Padoan model if it is to lead to accurate predictions for the RDF. In particular, the model predictions for $S^p_{2\parallel}$ and $S^p_{2\perp}$ must be scale-invariant in the dissipation range, in order to generate the well-known scale-invariant form of the RDF in this same range. However, we showed that the Pan \& Padoan model, and its modified forms that include the BIT dispersion theory, fail catastrophically in predicting scale-invariant solutions for  $S^p_{2\parallel}$ and $S^p_{2\perp}$ when $St=\mathcal{O}(1)$. These findings highlight the great difficulty in constructing fully closed, analytical predictions for the statistics of inertial particle relative motion in the dissipation range when $St=\mathcal{O}(1)$, where simple perturbation methods are of no use.

\section*{Acknowledgements}
We gratefully acknowledge Dr. Peter J. Ireland for providing the DNS data used in this paper.

\bibliographystyle{unsrt}
\bibliography{refs_co12}

\end{document}